\documentclass[aps,prl,reprint,superscriptaddress]{revtex4-2}
\usepackage[usenames,dvipsnames]{xcolor} 
\usepackage{amssymb,amsmath,bm,braket} 
\usepackage{graphicx} 
\usepackage[bookmarks=true,colorlinks,citecolor=blue,urlcolor=blue]{hyperref} 
\begin{document}

\title{Sine-square deformed mean-field theory}
\author{Masataka Kawano}
\email{masataka.kawano@tum.de}
\author{Chisa Hotta}
\affiliation{Department of Basic Science, University of Tokyo, Meguro-ku, Tokyo 153-8902, Japan}
\date{\today}

\begin{abstract}
We develop a theory that accurately evaluates quantum phases with any large-scale emergent structures 
including incommensurate density waves or topological textures without {\it a priori} knowing their periodicity. 
We spatially deform a real-space mean-field Hamiltonian on a finite-size cluster
using a sine-squared envelope function with zero energy at system edges.
The wave functions become insensitive to the misfit of the lattice and ordering periods.
We successfully extract the ordering wave vectors by our deformed Fourier transformation, 
updating the previous results for hole-doped and spin-orbit coupled Mott insulators. 
The method further enables the evaluation of a charge gap beyond the mean-field level. 
\end{abstract}
\maketitle

\textit{Introduction.}
How to detect and describe a variety of ordered quantum phases having a long period
or a large-scale structure in crystalline solids is one of the major theoretical challenges.
Over the last decades, the condensed matter field has focused on these
emergent phases with spatially extended periods that exhibit extraordinary properties due to topology.
Examples include magnetic skyrmions with swirling spin structures in two dimensions \cite{robler2006nature,muhlbauer2009science,yu2010nature,heinze2011natphys}
and a chiral soliton lattice in quasi one-dimension (1D) \cite{dzyaloshinskii1964spj,dzyaloshinskii1965spj,moriya1982ssc,miyadai1983jpsj,togawa2012prl}.
These incommensurate spin textures originate from spin-orbit coupling (SOC) or competing local interactions
due to geometric frustration, and serve as the potential platform of next-generation memory devices and topological transport phenomena \cite{togawa2016jpsj,nagaosa2013natnano}.
More generally, the electron/hole doping often induces charge-density-wave (CDW) and spin-density-wave (SDW) orders,
which are considered as the key to understanding the high-temperature superconductivity \cite{bednorz1986zpb,wu2011nature}
and colossal magnetoresistance \cite{jin1994science,uehara1999nature}. 
Moir\'{e} graphene has a long-wavelength electronic state engineered by a misfit of two layers,
and there is an urgent demand to understand the origin of its unconventional superconductivity \cite{cao2018nature}.
\par
The theoretical difficulty in dealing with incommensurate orders in {\it quantum} many-body systems
lies in that they have exceptionally large unit cells whose periods depend sensitively on the parameters of the Hamiltonian.
In methods that rely on the reciprocal lattice representation such as a mean-field calculation,
the candidate period needs to be known \textit{apriori}.
Cluster-based methods such as the cluster dynamical mean-field theory \cite{metzner1989prl,georges1992prb,lichtenstein2000prb,kotliar2001prl}
or the cluster perturbation theory \cite{senechal2000prl,senechal2002prb}
have crucial problems, since the orders with periods that match the cluster size are energetically favored by accident.
In calculating quantum many-body states on finite clusters, such as exact diagonalization,
density matrix renormalization group (DMRG) \cite{white1992prl,white1993prb,schollwock2011ap},
and tensor network methods \cite{nishio2004arxiv,jordan2008prl,orus2014ap},
the results often depend sensitively on the boundary conditions \cite{shibata2011prb}, 
size and shape of the clusters\cite{hotta2012prb,hotta2013prb}. 
This becomes serious in dealing with incommensurate orders since the mismatch of the period between an order and
size of the clusters excludes the true orders from the lowest-energy-state candidates.
This can also be a problem for variational Monte Carlo methods \cite{gros1987prb,gros1988prb,giamarchi1991prb,dev1992prb,sorella2005prb}
and quantum Monte Carlo (QMC) methods \cite{suzuki1976ptp,foulkes2001rmp}
which can deal with large system size, when several orderings compete with each other. 
Notice that such a problem does not occur for classical systems, since 
the thermal equilibrium state does not require a coherence throughout the system unlike the wave functions in quantum phases. 
In fact, the open or free-boundary conditions can offer a proper evaluation of large-scale skyrmions in classical Monte Carlo simulations \cite{okubo2021prl}.
\par
In this Letter, we develop a sine-square deformed mean-field theory (SSDMF),
a protocol that can accurately evaluate the lowest-energy quantum state having an incommensurate order
without knowing it {\it apriori}.
The sine-square deformation (SSD) modifies a Hamiltonian $\mathcal{H}$ by an envelope function $f_{\mathrm{SSD}}(\bm{r})$ as 
${\mathcal{H}}_{\mathrm{SSD}}=\int d\bm{r}f_{\mathrm{SSD}}(\bm{r}){\mathcal{H}}(\bm{r})$, where $f_{\mathrm{SSD}}(\bm{r})$ smoothly scales down 
from unity at the system center ($\bm{r}=0$) towards zero at the edges 
\cite{gendiar2009ptp,hikihara2011prb,gendiar2011pra}. 
By now it is widely used for quantum many-body ground states, dynamics, finite temperature, and for field theories
\cite{tada2015mpl,okunishi2015jpa,yonaga2015jpsj,okumishi2016ptep,wen2016prb,morita2016prb,morita2016jpsj,hotta2018prb,kadosawa2020jpsj,hotta2021arxiv,takuma2021arxiv}. 
Deforming the whole terms of the Hamiltonian is known to keep the nature of the Hamiltonian unchanged \cite{katsura2011jpa,maruyama2011prb,katsura2012jpa,hikihara2013pra},
since it works as a real-space renormalization \cite{hotta2013prb}.
At the same time, it reduces the boundary-effect and finite-size effects since the edges have zero-energies
\cite{shibata2011prb,hotta2012prb,nishimoto2013ncom,hotta2013prb}. 
When applying the SSD using the DMRG, even commensurate-incommensurate quantum phase transitions are very accurately evaluated \cite{hotta2012prb}. 
Unfortunately, their wave vectors were inaccessible for two reasons;
the many-body wave functions do not provide meaningful spatial profile for 
the one-body quantities as well as two-point correlations with the SSD. 
Even when the spatial structures were available, 
the standard Fourier transformation is not applied to SSD systems which break the translational symmetry. 
We resolve this issue by developing a dubbed SSDMF and by introducing a deformed Fourier transformation. 
The ordering wave vector is extracted accurately 
even when the system size is much smaller compared to the period of the order,
and the result does not change in extrapolating it to infinite system size. 

\textit{SSDMF.}
Let us consider a generalized Hubbard Hamiltonian, 
$\hat{\mathcal{H}}=\hat{\mathcal{H}}_{0}+\hat{\mathcal{H}}_{U}$ with 
\begin{align}
\hat{\mathcal{H}}_{0}
&=
\sum_{i,j=1}^{N}
\sum_{\sigma=\uparrow,\downarrow}
f\left(
\frac{\bm{r}_{i}+\bm{r}_{j}}{2}
\right)
t_{i,j}^{\sigma}
\hat{c}_{i,\sigma}^{\dagger}
\hat{c}_{j,\sigma}
-
\mu
\sum_{i=1}^{N}
f(\bm{r}_{i})
\hat{n}_{i}
,
\label{eq:H0}
\\
\hat{\mathcal{H}}_{U}
&=
U
\sum_{i=1}^{N}
f(\bm{r}_{i})
\hat{n}_{i,\uparrow}
\hat{n}_{i,\downarrow}
,
\label{eq:HU}
\end{align}
where $N$ is the number of sites ($N=L\times L$ for the square lattice),
$t_{i,j}^{\sigma}=(t_{j,i}^{\sigma})^{*}$ is the directed transfer integral from 
site $j$ to $i$ with spin $\sigma$, 
$\mu$ is the chemical potential, 
$\hat{c}_{i,\sigma}^{\dagger}/\hat{c}_{i,\sigma}$ is the creation/annihilation operator of a fermion 
with particle density $\hat{n}_{i,\sigma}=\hat{c}_{i,\sigma}^{\dagger}\hat{c}_{i,\sigma}$ and $\hat{n}_{i}=\sum_{\sigma}\hat{n}_{i,\sigma}$.
We take the origin of the positional vector $\bm{r}_{i}$ at the center of the system. 
The envelop function for a periodic boundary condition (PBC) is $f(\bm{r})=f_{\mathrm{PBC}}(\bm{r})=1$ and for SSD is \cite{hotta2013prb},
\begin{equation}
f(\bm{r})
=
f_{\mathrm{SSD}}(\bm{r})
=
\frac{1}{2}
\left\{
1
+
\cos
\left(
\frac{\pi|\bm{r}|}{R}
\right)
\right\}, 
\label{eq:fssd}
\end{equation}
with a radius $R=R_0+1/2$ where $R_{0}$ is the distance of the farthest site from the origin. 
This function smoothly modifies the energy scale of the Hamiltonian 
from $f_{\mathrm{SSD}}(\bm{r})\sim1$ at the center to $\sim 0$ at the edges. 
\par
Originally the eigenstates are characterized by their wavenumber $k$.
However, the deformation-induced term $\propto(1/2-f_{\mathrm{SSD}}(\bm{r}))$
introduces a moderate scattering between these $k$-indexed states 
and generates a basis of localized wave-packets \cite{hotta2012prb,hotta2013prb}. 
Each wave packet has a major contribution to the eigenstate of $\hat{\mathcal{H}}$
with the eigenenergy that matches the energy scale of its location. 
Since wave packets are localized, the finite size effect is suppressed 
and the excitation energy scales with system size $L$ as $\propto 1/L^2$ \cite{hotta2013prb}.
The wave packets at the edges have zero energy and serve as a reservoir,
while a set of states that contribute to the center reproduce intrinsic ground state properties of the original Hamiltonian.  
The number of particles is automatically adjusted between them for a given value of $\mu/t_{i,j}^{\sigma}$ and $U/t_{i,j}^{\sigma}$.
\par
Our mean-field Hamiltonian for the SSD is given by 
\begin{equation}
\hat{\mathcal{H}}_{\mathrm{MF}}
=
\hat{\mathcal{H}}_{0}
+
U
\sum_{i=1}^{N}
f(\bm{r}_{i})
\left(
\frac{\braket{\hat{n}_{i}}}{2}
\hat{n}_{i}
-
2
\braket{\hat{\bm{S}}_{i}}
\cdot
\hat{\bm{S}}_{i}
\right) +
E_{c}
,
\label{eq:Hmf}
\end{equation}
with $E_{c}\!=\!-U\sum_{i=1}^{N}f(\bm{r}_{i})\{\braket{\hat{n}_{i}}^{2}\!/4-\braket{\hat{\bm{S}}_{i}}^{2}\}$. 
Here, $\braket{\hat{n}_{i}}$ and $\braket{\hat{\bm{S}}_{i}}$
are the order parameters representing the particle and spin densities, respectively, 
which are determined self-consistently. 
The spin operator is defined as $\hat{\bm{S}}_{i}=(\hat{S}_{i}^{x},\hat{S}_{i}^{y},\hat{S}_{i}^{z})$
with $\hat{S}_{i}^{\mu}=(1/2)\sum_{\sigma,\sigma'}\hat{c}_{i,\sigma}^{\dagger}\tau_{\sigma,\sigma'}^{\mu}\hat{c}_{i,\sigma'}$,
where $\tau^{\mu}$ ($\mu=x,y,z$) is the Pauli matrix. 
The mean fields serve as internal SSD fields 
and are self-consistently determined by minimizing the energy.

\begin{figure}[t]
\includegraphics[width=85mm]{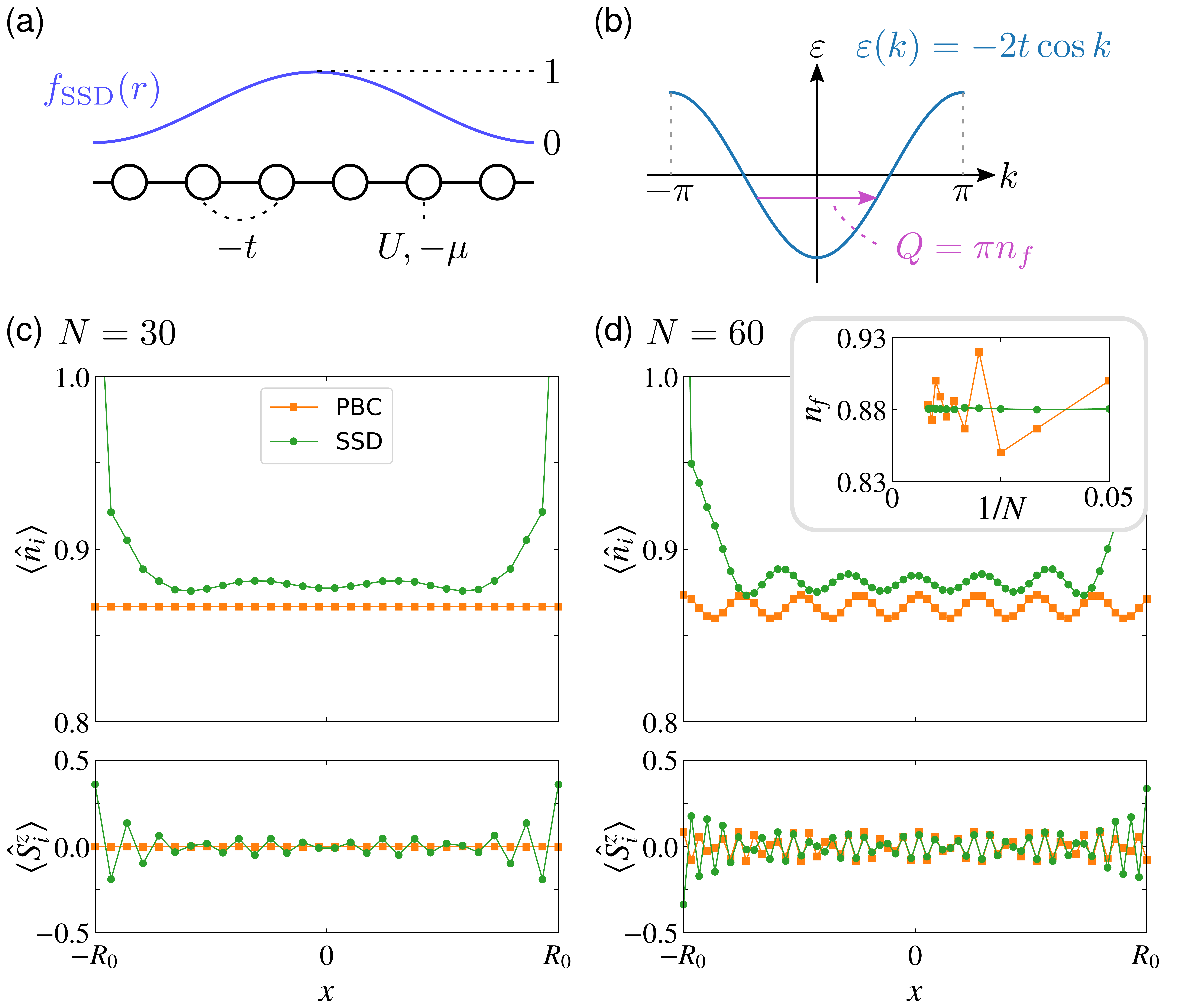}
\caption{(a) Schematic illustration of the 1D Hubbard model with the SSD.
(b) Noninteracting band $\varepsilon(k)=-2\cos k$ and nesting vector $Q=\pi n_{f}$.
The hole doping $n_{f}<1$ shifts the nesting vector off $\pi$. 
(c,d) Spatial distribution of the particle density and spin density
at $U=2.0$, $\mu=0.5$ with $N=30$ and $60$. 
The inset in (d) shows the size dependence of the particle density $n_{f}$ 
obtained by Eq.(\ref{eq:nf}). }
\label{fig1}
\end{figure}
\textit{Hole-doped 1D Hubbard model.}
We first consider the hole-doped 1D Hubbard model with the PBC and SSD (Fig.~\ref{fig1}(a)) 
by setting $t_{i,j}^{\sigma}=-1$ uniform and only between nearest-neighbor sites. 
The hole doping shifts the nesting vector $Q \ne \pi$ as shown in Fig.~\ref{fig1}(b),
and yields an incommensurate SDW of wavenumber $Q$ and CDW of $2Q$ at the mean-field level \cite{ichioka2001jpsj}. 
The coexistence is allowed since the latter is the higher harmonics of the former \cite{ichioka2001jpsj,ono2000jpsj}.
Although the Bethe-ansatz solutions indicate the absence of any long-range orders \cite{bethe1931zp,lieb1968prl},
the mean-field approximation can still capture the dominant wavenumber that should appear in its correlation functions. 
It also provides a good description for actual materials where a finite inter-chain coupling is relevant.
We will show how the SSDMF eliminates the finite-size effects arising from the long-period CDW/SDW order.
\par
We consider $U=2.0$ and $\mu=0.5$ which is less than half-filling, 
and set the initial value of the mean fields to those with 
a U(1) spin-rotational symmetry about the $z$-axis, which can describe the CDW/SDW order.
Figures~\ref{fig1}(c) and \ref{fig1}(d) show the spatial distribution of 
particle and spin densities for $N=30$ and $N=60$, respectively. 
The PBC system does not exhibit any spatial structure for $N=30$,
while the SSD system shows a clear modulation of the particle density by optimizing 
their particle and spin densities using the edges as a reservoir.
For $N\geq60$, both the SSD and PBC show features of CDW and SDW orderings. 
To quantitatively evaluate the results, 
we extract the particle density by the following formula \cite{gendiar2009ptp}: 
\begin{equation}
n_{f}
=
\frac{\sum_{i=1}^{N}f(x_{i})\braket{\hat{n}_{i}}}{\sum_{i=1}^{N}f(x_{i})}, 
\label{eq:nf}
\end{equation}
which is reduced to the standard average particle density for the PBC. 
Indeed, the PBC still has a translationally symmetric wave function
for periods larger than the lattice spacings in the mean-field solution. 
However, for the SSD such a symmetry largely deteriorates at the edges,
and simply averaging the whole particle density gives a physically meaningless result. 
Equation (\ref{eq:nf}) extracts the contribution near the center by
smoothly suppressing the contribution closer to the edges. 
As shown in the inset of Fig.~\ref{fig1}(d), the extracted $n_{f}$ for the SSD 
is almost $N$-independent; we find the value at the thermodynamic limit $n_{f}\simeq0.88$ already at $N=20$.
Whereas the ones obtained by the PBC show a large oscillation even at around $N\sim 100$. 
\par
\begin{figure}[t]
\includegraphics[width=85mm]{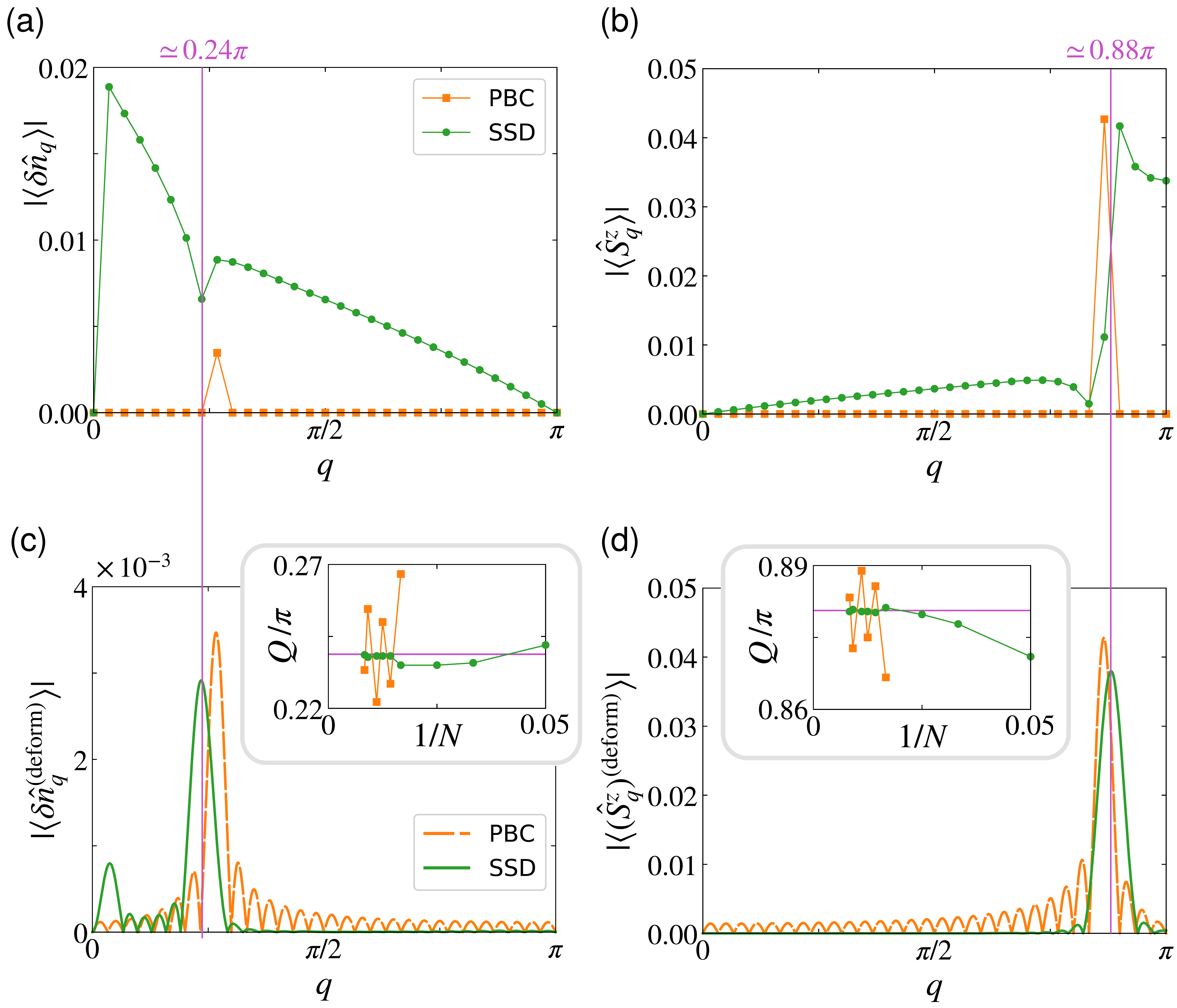}
\caption{(a,b) Absolute value of the standard Fourier component of the (a) particle-density deviation $|\braket{\delta\hat{n}_{q}}|$
and (b) $z$-component of the spin density $|\braket{\hat{S}_{q}^{z}}|$ as the function of wave number $q$ 
for the hole-doped 1D Hubbard model with $N=60$.
The purple line denotes the ordering wave number in the thermodynamic limit.
In the SSD, there are large background contributions mainly from the edge sites. 
(c,d) Deformed Fourier component 
$|\braket{\delta\hat{n}_{q}^{(\mathrm{deform})}}|$ and $|\braket{(\hat{S}_{q}^{z})^{(\mathrm{deform})}}|$ 
obtained by Eq. (\ref{eq:deformed-Fourier}). 
Although $q$ is taken as continuous in equivalently adopting Eq.(\ref{eq:deformed-Fourier}) to PBC and SSD,
for PBC it is meaningful for $q=2\pi/N \times $(integer) which are peak positions.
Insets: $1/N$-dependence of the ordering wave number $Q$ extracted from the peak position in the main panels.}
\label{fig2}
\end{figure}

Next, we show that the ordering wavenumbers can be accurately evaluated. 
We subtract the uniform component from the particle density as 
$\delta\hat{n}_{i}=\hat{n}_{i}-n_{f}$. 
We first demonstrate that the standard Fourier transformation,  
$\hat{o}_{q}=(1/N)\sum_{j=1}^{N}\hat{o}_{j}\mathrm{e}^{-iqx_{j}}$ 
for local operator $\hat{o}_{i}$ and wavenumber $q=2n\pi/N$ ($n\in\mathbb{Z}$),
does not give meaningful results.
In Figs.~\ref{fig2}(a) and \ref{fig2}(b) these Fourier components,
$|\braket{\delta\hat{n}_{q}}|$ and $|\braket{\hat{S}_{q}^{z}}|$, are shown as the function of $q$.
Those for the SSD have the significantly large contribution from the edge,
and although we find spikes in the SSD data, their positions are not equal to the PBC ones nor to those at the thermodynamic limit,
making it hard to understand their physical implication.
To resolve this issue, we define a deformed Fourier transformation as
\begin{equation}
\hat{o}_{q}^{(\mathrm{deform})}
=
\frac
{\sum_{j=1}^{N}f(x_{j})\hat{o}_{j}\mathrm{e}^{-iqx_{j}}}
{\sum_{j=1}^{N}f(x_{j})}
,
\label{eq:deformed-Fourier}
\end{equation}
and we take $q$ as a continuous variable. 
Equation (\ref{eq:deformed-Fourier}) is regarded as a generalization of the Fourier transformation
applied to systems with spatial variation, and has several advantages over the standard
transformation. 
First, the average particle density corresponds to the $q=0$ component as 
$\braket{\hat{n}_{q=0}^{(\mathrm{deform})}}=n_{f}$,
and is reduced to the standard Fourier transformation when we use $f_{\mathrm{PBC}}(x_{j})$. 
Second, the finite size effect that usually appears due to the discretization of $q$ is formally smeared out. 
Figures~\ref{fig2}(c) and \ref{fig2}(d) show $|\braket{\delta\hat{n}_{q}^{(\mathrm{deform})}}|$
and $|\braket{(\hat{S}_{q}^{z})^{(\mathrm{deform})}}|$ as the function of continuous $q$. 
For the PBC, although the peak position does not change from 
Figs.~\ref{fig2}(a) and \ref{fig2}(b), some additional peaks appear. 
For the SSD, our transformation completely suppresses the background contribution 
and the clear peak structure appears, whose position is close to that of the thermodynamic limit.
The insets in Figs.~\ref{fig2}(c) and \ref{fig2}(d) show 
the size dependence of the wave number extracted from the peak. 
Again the ones from the SSD have no significant size dependence, 
and the peak position already reaches the thermodynamic limit at $N\geq60$.
Therefore, the SSDMF can efficiently describe incommensurate orders even when the system size is small,
and more importantly, there is no artifact from the mismatch of the period between the order and lattice; 
by minimizing the total energy of the system 
the excess/deficient particles and spins are automatically 
removed from the system center using the almost zero-energy edge sites as a reservoir.
%
\par
\begin{figure}[t]
\includegraphics[width=85mm]{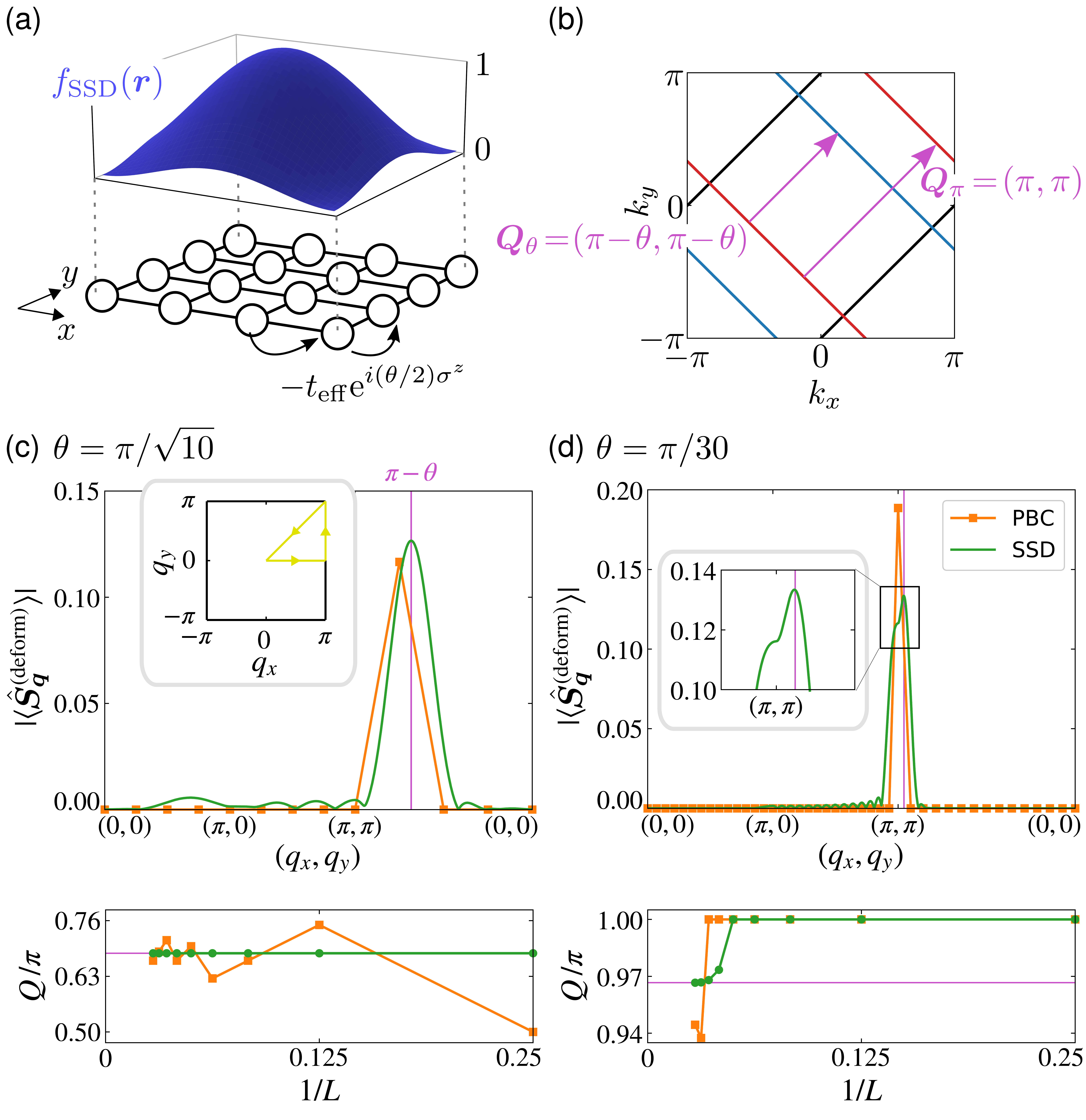}
\caption{(a) Schematic illustration of the SSD square-lattice Hubbard model with SOC.
(b) Fermi surfaces of the noninteracting up (red line) and down (blue line) spin 
bands with nesting vectors $\bm{Q}_{\theta}=(\pi-\theta,\pi-\theta)$ and $\bm{Q}_{\pi}=(\pi,\pi)$, 
and that of the doubly degenerate bands (black). 
Finite SOC $\lambda=t_{\mathrm{eff}}\sin(\theta/2)$ induces the incommensurate spiral order 
characterized by $|\braket{\hat{\bm{S}}_{q}^{(\mathrm{deform})}}|$
as the function of $\bm{q}=(q_{x},q_{y})$.
The results for $U=2.0$, $\mu=1.0$ are shown along the symmetric line in the Brillouin zone (inset of (c)), 
for (c) $L=8$, $\theta=\pi/\sqrt{10}$ and (d) $L=28$, $\theta=\pi/30$.
The purple dashed line denotes the ordering wave number at the thermodynamic limit.
The bottom panels show the size dependence of the peak position $\bm{Q}=(Q,Q)$.
The $(\pi,\pi)$ peak for PBC is a size effect that approaches the true $\bm{Q}$ at $N\rightarrow\infty$.
}
\label{fig3}
\end{figure}

\textit{Square-lattice Hubbard model with SOC.}
We consider a square-lattice Hubbard model at half-filling 
with SOC which manifests in a spin-dependent transfer integral $\lambda$ 
that rotates the spin about the $z$-axis, as shown in Fig.~\ref{fig3}(a); 
the nearest-neighbor transfer integrals are 
$t_{i,j}^{\uparrow}=-t-i\lambda$ and $t_{i,j}^{\downarrow}=-t+i\lambda$ when propagating in the $+x$ or $+y$ direction.
This form is realized by the combination of Rashba and Dresselhaus SOC with equal amplitudes, 
and is known to host persistent spin helix states \cite{bernevig2006prl,john2017rmp}.
It is convenient to rewrite the first term in Eq. (\ref{eq:H0}) using an SU(2) gauge field as
\begin{equation}
\sum_{\sigma}
t_{i,j}^{\sigma}
\hat{c}_{i,\sigma}^{\dagger}
\hat{c}_{j,\sigma}
=
-t_{\mathrm{eff}}
\hat{\bm{c}}_{i}^{\dagger}
\mathrm{e}^{i(\theta/2)\sigma^{z}}
\hat{\bm{c}}_{j}
,
\label{eq:SOC}
\end{equation}
where $\hat{\bm{c}}_{i}=(\hat{c}_{i,\uparrow},\hat{c}_{i,\downarrow})^{T}$,
$t_{\mathrm{eff}}=\sqrt{t^{2}+\lambda^{2}}$, and $\theta=2\mathrm{arctan}(\lambda/t)$.
We take $t_{\mathrm{eff}}=1$ in the following.
\par
The Fermi surfaces of the noninteracting bands are shown in Fig.~\ref{fig3}(b). 
There are two vectors, 
$\bm{Q}_{\theta}=(\pi-\theta,\pi-\theta)$ and $\bm{Q}_{\pi}=(\pi,\pi)$,
that perfectly nest the Fermi-surfaces and drive the system to an insulator at infinitesimal $U$.
This Hamiltonian provides a prototype example
that the numerical calculations mislead the conclusions about the incommensurate orderings:
a recent mean-field study \cite{park2020prr} reports the existence of CDW and incommensurate SDW orders.
However, such possibility is excluded by the following analytical argument. 
\par
Let us perform a local unitary transformation using the operator 
$\hat{\mathcal{U}}=\bigotimes_{i=1}^{N}\exp(-i\theta(x_{i}+y_{i})\hat{S}_{i}^{z})$
that rotates the spin quantization axis. 
SOC in Eq. (\ref{eq:SOC}) disappears as 
$\hat{\mathcal{U}}\hat{\bm{c}}_{i}^{\dagger}\mathrm{e}^{i(\theta/2)\sigma^{z}}\hat{\bm{c}}_{j}\hat{\mathcal{U}}^{\dagger}
=\sum_{\sigma}\hat{c}_{i,\sigma}^{\dagger}\hat{c}_{j,\sigma}$, 
and the Hamiltonian is transformed to the SU(2) symmetric Hubbard model,
which shows a standard antiferromagnetic ordering.
In addition, there is a rigorous proof that the CDW order is absent \cite{kubo1990prb}.
By transforming back the antiferromagnetic ordering of the rotating frame 
to the original frame, we obtain a spiral ordering. 
Therefore, the reported CDW/SDW orderings \cite{park2020prr} are determinably the artifact of the standard mean-field framework.
\par
Our SSDMF correctly captures the spiral-ordered ground state and its periodicity. 
We take $U=2.0$, $\mu=1.0$ which is half-filling, and an SOC with $\theta=\pi/\sqrt{10}$ and $\theta=\pi/30$. 
We prepare the initial mean fields that allow the spiral order in the $xy$-plane,
which is characterized by the wave vector $\bm{Q}_{\theta}$. 
Figures~\ref{fig3}(c) and \ref{fig3}(d) show $|\braket{\hat{\bm{S}}_{\bm{q}}^{(\mathrm{deform})}}|$ 
and the $1/L$-dependence of peak positions $\bm{Q}=(Q,Q)$. 
Although $Q/\pi$ takes an irrational number for $\theta=\pi/\sqrt{10}$, 
the SSDMF gives that exact value for the system as small as $L=4$. 
When $\theta=\pi/30$, the expected spiral order has a long periodicity of size $60\times60$; 
the method even succeeds in accurately extracting this period, 
$\bm{Q}=(0.967\pi,0.967\pi)$ (unchanged up to $N\rightarrow\infty$), 
which is much longer than the size of the cluster $L \sim 24$. 
We checked several different initial states, finding that they converge to the spirals.
\begin{figure}[t]
\includegraphics[width=85mm]{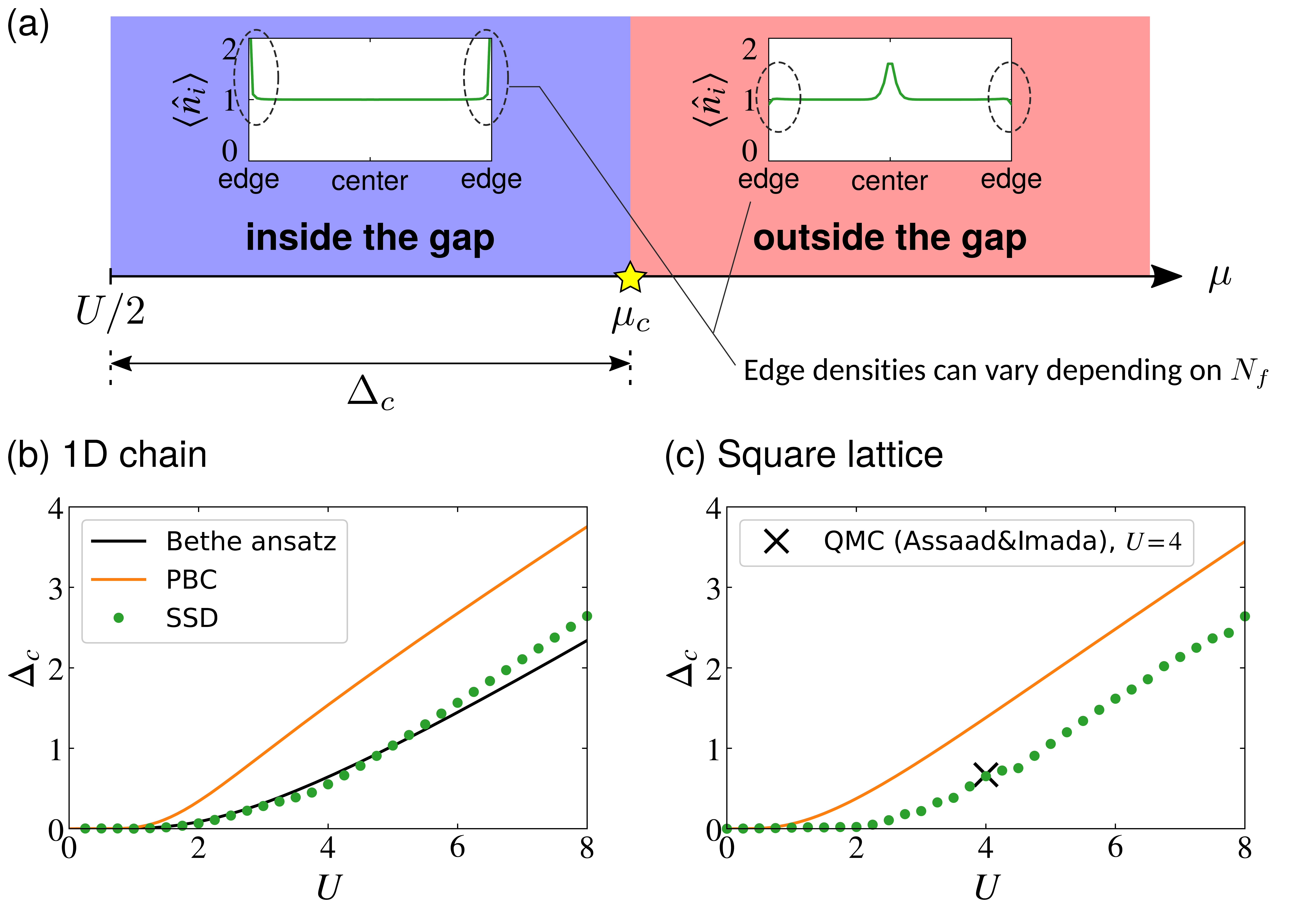}
\caption{(a) Calculation scheme of the charge gap in the SSDMF. 
The charge density at the center is determined by $\mu$, while varying $N_f$ will vary the charge density at the edges for all cases. 
We plot the charge density obtained by the SSD calculation for 
$N_f$ close to the ideal filling factor determined by $\mu$ close to $\mu_c$. 
The two panels have the same $N_f$, while $\mu$ is varied; 
at $\mu \le \mu_c$ i.e. $|\mu-U/2|<\Delta_{c}$ (inside the Mott gap), 
the particles exceeding the filling are pushed to the system edge. 
At $\mu>\mu_{c}$, the particles are trapped (doped) a the system center 
and the critical value determines the charge gap $\mu_c-U/2=\Delta_{c}$.
(b,c) Charge gap as the function of $U$ for the (b) 1D Hubbard model with $N=120$ and (c) square-lattice Hubbard model with $L=40$.
The SSDMF well reproduces the exact results, even at the mean-field level.}
\label{fig4}
\end{figure}

\textit{Charge gap.}
Finally, we show that the SSDMF yields an accurate charge gap of the bulk Hubbard model 
beyond the mean-field level. 
Since SSD systems have a gapless spectrum by construction due to zero-energy edge modes, 
the standard formula, 
$\Delta_{c}=(E(N+2)+E(N-2)-2E(N))/4$,
does not give the appropriate charge gap, 
where $E(N_{f})$ is the $N_{f}$-particle ground-state energy. 
For the SSD, the intrinsic filling factor observed as a particle density at the system center
is determined not by $N_{f}$ but by $\mu$, 
which is numerically proven in the 1D Hubbard model using DMRG by comparing with the Bethe-ansatz solution \cite{hotta2012prb}.
Varying $N_f$ does not change the center particle density if $\mu$ is unchanged.
For $\mu$ in the gapless region, if $N_{f}$ is larger/smaller than the value expected,
the excess particles/holes concentrate on edges that are regarded as baths.
When $\mu$ is inside the gap, the center particle density remains constant even when we change $\mu$ or $N_{f}$.
At the point where $\mu\ge \mu_c$ starts to go outside the gap,
the particles start to be doped at around the center site (see Fig.~\ref{fig4}(a)).
\par
In our SSDMF by varying $\mu$ for an appropriate choice of $N_f$(not strict),
and measuring the critical value $\mu_{c}$, the charge gap is evaluated as $\Delta_{c}=\mu_{c}-U/2$,
where $U/2$ is the chemical potential at half-filling. 
The initial mean fields should be chosen carefully such that the final form of mean fields 
is symmetric about the system center. 
Figures~\ref{fig4}(b) and \ref{fig4}(c) show the charge gap for the 1D chain and square lattice. 
Here the SOC is turned off while turning it on does not change the result as proven by the local unitary transformation.
The PBC ones give the standard mean-field solutions, whereas our SSDMF is close to the 
exact Bethe ansatz solutions \cite{bethe1931zp,lieb1968prl} and highly accurate QMC results \cite{furukawa1991jpsj,assaad1996prl}.
The correlation effect beyond the mean field approximation is seemingly erased by the SSD.

\textit{Conclusion.} 
We proposed the SSDMF and the deformed Fourier transformation 
that accurately characterize the incommensurate orderings in an unbiased manner without knowing their spatial periods {\it apriori}.
The SSDMF also extracts the charge gap in high quality very close to the values of the
Bethe ansatz and QMC results in the bulk limit. 
Therefore, we conclude that the method provides decisive conclusions to 
many topics of modern condensed matter physics related to topology, 
where often the contradictory results from different numerical solvers previously made the issue controversial.

\textit{Acknowledgements.}
M. K. was supported by a Grant-in-Aid for JSPS Research Fellow (Grant No. 19J22468).
C.H. was supported by JSPS KAKENHI Grants No. JP17K05533, No.JP18H01173, No.JP21H05191, and No.JP21K03440 from the Ministry of Education, Science, Sports and Culture of Japan.

\bibliography{biblio}

\begin{thebibliography}{68}%
\makeatletter
\providecommand \@ifxundefined [1]{%
 \@ifx{#1\undefined}
}%
\providecommand \@ifnum [1]{%
 \ifnum #1\expandafter \@firstoftwo
 \else \expandafter \@secondoftwo
 \fi
}%
\providecommand \@ifx [1]{%
 \ifx #1\expandafter \@firstoftwo
 \else \expandafter \@secondoftwo
 \fi
}%
\providecommand \natexlab [1]{#1}%
\providecommand \enquote  [1]{``#1''}%
\providecommand \bibnamefont  [1]{#1}%
\providecommand \bibfnamefont [1]{#1}%
\providecommand \citenamefont [1]{#1}%
\providecommand \href@noop [0]{\@secondoftwo}%
\providecommand \href [0]{\begingroup \@sanitize@url \@href}%
\providecommand \@href[1]{\@@startlink{#1}\@@href}%
\providecommand \@@href[1]{\endgroup#1\@@endlink}%
\providecommand \@sanitize@url [0]{\catcode `\\12\catcode `\$12\catcode
  `\&12\catcode `\#12\catcode `\^12\catcode `\_12\catcode `\%12\relax}%
\providecommand \@@startlink[1]{}%
\providecommand \@@endlink[0]{}%
\providecommand \url  [0]{\begingroup\@sanitize@url \@url }%
\providecommand \@url [1]{\endgroup\@href {#1}{\urlprefix }}%
\providecommand \urlprefix  [0]{URL }%
\providecommand \Eprint [0]{\href }%
\providecommand \doibase [0]{https://doi.org/}%
\providecommand \selectlanguage [0]{\@gobble}%
\providecommand \bibinfo  [0]{\@secondoftwo}%
\providecommand \bibfield  [0]{\@secondoftwo}%
\providecommand \translation [1]{[#1]}%
\providecommand \BibitemOpen [0]{}%
\providecommand \bibitemStop [0]{}%
\providecommand \bibitemNoStop [0]{.\EOS\space}%
\providecommand \EOS [0]{\spacefactor3000\relax}%
\providecommand \BibitemShut  [1]{\csname bibitem#1\endcsname}%
\let\auto@bib@innerbib\@empty
\bibitem [{\citenamefont {R\"{o}{\ss}ler}\ \emph {et~al.}(2006)\citenamefont
  {R\"{o}{\ss}ler}, \citenamefont {Bogdanov},\ and\ \citenamefont
  {Pfleiderer}}]{robler2006nature}%
  \BibitemOpen
  \bibfield  {author} {\bibinfo {author} {\bibfnamefont {U.~K.}\ \bibnamefont
  {R\"{o}{\ss}ler}}, \bibinfo {author} {\bibfnamefont {A.~N.}\ \bibnamefont
  {Bogdanov}},\ and\ \bibinfo {author} {\bibfnamefont {C.}~\bibnamefont
  {Pfleiderer}},\ }\bibfield  {title} {\bibinfo {title} {{Spontaneous skyrmion
  ground states in magnetic metals}},\ }\href
  {https://doi.org/10.1038/nature05056} {\bibfield  {journal} {\bibinfo
  {journal} {Nature}\ }\textbf {\bibinfo {volume} {442}},\ \bibinfo {pages}
  {797} (\bibinfo {year} {2006})}\BibitemShut {NoStop}%
\bibitem [{\citenamefont {M\"{u}hlbauer}\ \emph {et~al.}(2009)\citenamefont
  {M\"{u}hlbauer}, \citenamefont {Binz}, \citenamefont {Jonietz}, \citenamefont
  {Pfleiderer}, \citenamefont {Rosch}, \citenamefont {Neubauer}, \citenamefont
  {Georgii},\ and\ \citenamefont {B\"{o}ni}}]{muhlbauer2009science}%
  \BibitemOpen
  \bibfield  {author} {\bibinfo {author} {\bibfnamefont {S.}~\bibnamefont
  {M\"{u}hlbauer}}, \bibinfo {author} {\bibfnamefont {B.}~\bibnamefont {Binz}},
  \bibinfo {author} {\bibfnamefont {F.}~\bibnamefont {Jonietz}}, \bibinfo
  {author} {\bibfnamefont {C.}~\bibnamefont {Pfleiderer}}, \bibinfo {author}
  {\bibfnamefont {A.}~\bibnamefont {Rosch}}, \bibinfo {author} {\bibfnamefont
  {A.}~\bibnamefont {Neubauer}}, \bibinfo {author} {\bibfnamefont
  {R.}~\bibnamefont {Georgii}},\ and\ \bibinfo {author} {\bibfnamefont
  {P.}~\bibnamefont {B\"{o}ni}},\ }\bibfield  {title} {\bibinfo {title}
  {{Skyrmion Lattice in a Chiral Magnet}},\ }\href
  {http://science.sciencemag.org/content/323/5916/915.abstract} {\bibfield
  {journal} {\bibinfo  {journal} {Science}\ }\textbf {\bibinfo {volume}
  {323}},\ \bibinfo {pages} {915} (\bibinfo {year} {2009})}\BibitemShut
  {NoStop}%
\bibitem [{\citenamefont {Yu}\ \emph {et~al.}(2010)\citenamefont {Yu},
  \citenamefont {Onose}, \citenamefont {Kanazawa}, \citenamefont {Park},
  \citenamefont {Han}, \citenamefont {Matsui}, \citenamefont {Nagaosa},\ and\
  \citenamefont {Tokura}}]{yu2010nature}%
  \BibitemOpen
  \bibfield  {author} {\bibinfo {author} {\bibfnamefont {X.~Z.}\ \bibnamefont
  {Yu}}, \bibinfo {author} {\bibfnamefont {Y.}~\bibnamefont {Onose}}, \bibinfo
  {author} {\bibfnamefont {N.}~\bibnamefont {Kanazawa}}, \bibinfo {author}
  {\bibfnamefont {J.~H.}\ \bibnamefont {Park}}, \bibinfo {author}
  {\bibfnamefont {J.~H.}\ \bibnamefont {Han}}, \bibinfo {author} {\bibfnamefont
  {Y.}~\bibnamefont {Matsui}}, \bibinfo {author} {\bibfnamefont
  {N.}~\bibnamefont {Nagaosa}},\ and\ \bibinfo {author} {\bibfnamefont
  {Y.}~\bibnamefont {Tokura}},\ }\bibfield  {title} {\bibinfo {title}
  {{Real-space observation of a two-dimensional skyrmion crystal}},\ }\href
  {https://doi.org/10.1038/nature09124} {\bibfield  {journal} {\bibinfo
  {journal} {Nature}\ }\textbf {\bibinfo {volume} {465}},\ \bibinfo {pages}
  {901} (\bibinfo {year} {2010})}\BibitemShut {NoStop}%
\bibitem [{\citenamefont {Heinze}\ \emph {et~al.}(2011)\citenamefont {Heinze},
  \citenamefont {von Bergmann}, \citenamefont {Menzel}, \citenamefont {Brede},
  \citenamefont {Kubetzka}, \citenamefont {Wiesendanger}, \citenamefont
  {Bihlmayer},\ and\ \citenamefont {Bl\"{u}gel}}]{heinze2011natphys}%
  \BibitemOpen
  \bibfield  {author} {\bibinfo {author} {\bibfnamefont {S.}~\bibnamefont
  {Heinze}}, \bibinfo {author} {\bibfnamefont {K.}~\bibnamefont {von
  Bergmann}}, \bibinfo {author} {\bibfnamefont {M.}~\bibnamefont {Menzel}},
  \bibinfo {author} {\bibfnamefont {J.}~\bibnamefont {Brede}}, \bibinfo
  {author} {\bibfnamefont {A.}~\bibnamefont {Kubetzka}}, \bibinfo {author}
  {\bibfnamefont {R.}~\bibnamefont {Wiesendanger}}, \bibinfo {author}
  {\bibfnamefont {G.}~\bibnamefont {Bihlmayer}},\ and\ \bibinfo {author}
  {\bibfnamefont {S.}~\bibnamefont {Bl\"{u}gel}},\ }\bibfield  {title}
  {\bibinfo {title} {{Spontaneous atomic-scale magnetic skyrmion lattice in two
  dimensions}},\ }\href {https://doi.org/10.1038/nphys2045} {\bibfield
  {journal} {\bibinfo  {journal} {Nat. Phys.}\ }\textbf {\bibinfo {volume}
  {7}},\ \bibinfo {pages} {713} (\bibinfo {year} {2011})}\BibitemShut {NoStop}%
\bibitem [{\citenamefont {Dzyaloshinskii}(1964)}]{dzyaloshinskii1964spj}%
  \BibitemOpen
  \bibfield  {author} {\bibinfo {author} {\bibfnamefont {I.}~\bibnamefont
  {Dzyaloshinskii}},\ }\bibfield  {title} {\bibinfo {title} {{Theory of
  helicoidal structures in antiferromagnets. I. Nonmetals}},\ }\href@noop {}
  {\bibfield  {journal} {\bibinfo  {journal} {Sov. Phys. JETP}\ }\textbf
  {\bibinfo {volume} {19}},\ \bibinfo {pages} {960} (\bibinfo {year}
  {1964})}\BibitemShut {NoStop}%
\bibitem [{\citenamefont {Dzyaloshinskii}(1965)}]{dzyaloshinskii1965spj}%
  \BibitemOpen
  \bibfield  {author} {\bibinfo {author} {\bibfnamefont {I.}~\bibnamefont
  {Dzyaloshinskii}},\ }\bibfield  {title} {\bibinfo {title} {{The theory of
  helicoidal structures in antiferromagnets. II. metals}},\ }\href@noop {}
  {\bibfield  {journal} {\bibinfo  {journal} {Sov. Phys. JETP}\ }\textbf
  {\bibinfo {volume} {20}},\ \bibinfo {pages} {223} (\bibinfo {year}
  {1965})}\BibitemShut {NoStop}%
\bibitem [{\citenamefont {Moriya}\ and\ \citenamefont
  {Miyadai}(1982)}]{moriya1982ssc}%
  \BibitemOpen
  \bibfield  {author} {\bibinfo {author} {\bibfnamefont {T.}~\bibnamefont
  {Moriya}}\ and\ \bibinfo {author} {\bibfnamefont {T.}~\bibnamefont
  {Miyadai}},\ }\bibfield  {title} {\bibinfo {title} {{Evidence for the helical
  spin structure due to antisymmetric exchange interaction in
  Cr$_{13}$NbS$_{2}$}},\ }\href
  {https://www.sciencedirect.com/science/article/pii/0038109882910067}
  {\bibfield  {journal} {\bibinfo  {journal} {Solid State Commun.}\ }\textbf
  {\bibinfo {volume} {42}},\ \bibinfo {pages} {209} (\bibinfo {year}
  {1982})}\BibitemShut {NoStop}%
\bibitem [{\citenamefont {Miyadai}\ \emph {et~al.}(1983)\citenamefont
  {Miyadai}, \citenamefont {Kikuchi}, \citenamefont {Kondo}, \citenamefont
  {Sakka}, \citenamefont {Arai},\ and\ \citenamefont
  {Ishikawa}}]{miyadai1983jpsj}%
  \BibitemOpen
  \bibfield  {author} {\bibinfo {author} {\bibfnamefont {T.}~\bibnamefont
  {Miyadai}}, \bibinfo {author} {\bibfnamefont {K.}~\bibnamefont {Kikuchi}},
  \bibinfo {author} {\bibfnamefont {H.}~\bibnamefont {Kondo}}, \bibinfo
  {author} {\bibfnamefont {S.}~\bibnamefont {Sakka}}, \bibinfo {author}
  {\bibfnamefont {M.}~\bibnamefont {Arai}},\ and\ \bibinfo {author}
  {\bibfnamefont {Y.}~\bibnamefont {Ishikawa}},\ }\bibfield  {title} {\bibinfo
  {title} {{Magnetic Properties of Cr$_{1/3}$NbS$_{2}$}},\ }\href
  {https://doi.org/10.1143/JPSJ.52.1394} {\bibfield  {journal} {\bibinfo
  {journal} {J. Phys. Soc. Jpn.}\ }\textbf {\bibinfo {volume} {52}},\ \bibinfo
  {pages} {1394} (\bibinfo {year} {1983})}\BibitemShut {NoStop}%
\bibitem [{\citenamefont {Togawa}\ \emph {et~al.}(2012)\citenamefont {Togawa},
  \citenamefont {Koyama}, \citenamefont {Takayanagi}, \citenamefont {Mori},
  \citenamefont {Kousaka}, \citenamefont {Akimitsu}, \citenamefont {Nishihara},
  \citenamefont {Inoue}, \citenamefont {Ovchinnikov},\ and\ \citenamefont
  {Kishine}}]{togawa2012prl}%
  \BibitemOpen
  \bibfield  {author} {\bibinfo {author} {\bibfnamefont {Y.}~\bibnamefont
  {Togawa}}, \bibinfo {author} {\bibfnamefont {T.}~\bibnamefont {Koyama}},
  \bibinfo {author} {\bibfnamefont {K.}~\bibnamefont {Takayanagi}}, \bibinfo
  {author} {\bibfnamefont {S.}~\bibnamefont {Mori}}, \bibinfo {author}
  {\bibfnamefont {Y.}~\bibnamefont {Kousaka}}, \bibinfo {author} {\bibfnamefont
  {J.}~\bibnamefont {Akimitsu}}, \bibinfo {author} {\bibfnamefont
  {S.}~\bibnamefont {Nishihara}}, \bibinfo {author} {\bibfnamefont
  {K.}~\bibnamefont {Inoue}}, \bibinfo {author} {\bibfnamefont {A.~S.}\
  \bibnamefont {Ovchinnikov}},\ and\ \bibinfo {author} {\bibfnamefont
  {J.}~\bibnamefont {Kishine}},\ }\bibfield  {title} {\bibinfo {title} {{Chiral
  Magnetic Soliton Lattice on a Chiral Helimagnet}},\ }\href
  {https://link.aps.org/doi/10.1103/PhysRevLett.108.107202} {\bibfield
  {journal} {\bibinfo  {journal} {Phys. Rev. Lett.}\ }\textbf {\bibinfo
  {volume} {108}},\ \bibinfo {pages} {107202} (\bibinfo {year}
  {2012})}\BibitemShut {NoStop}%
\bibitem [{\citenamefont {Togawa}\ \emph {et~al.}(2016)\citenamefont {Togawa},
  \citenamefont {Kousaka}, \citenamefont {Inoue},\ and\ \citenamefont
  {Kishine}}]{togawa2016jpsj}%
  \BibitemOpen
  \bibfield  {author} {\bibinfo {author} {\bibfnamefont {Y.}~\bibnamefont
  {Togawa}}, \bibinfo {author} {\bibfnamefont {Y.}~\bibnamefont {Kousaka}},
  \bibinfo {author} {\bibfnamefont {K.}~\bibnamefont {Inoue}},\ and\ \bibinfo
  {author} {\bibfnamefont {J.-i.}\ \bibnamefont {Kishine}},\ }\bibfield
  {title} {\bibinfo {title} {{Symmetry, Structure, and Dynamics of Monoaxial
  Chiral Magnets}},\ }\href {https://doi.org/10.7566/JPSJ.85.112001} {\bibfield
   {journal} {\bibinfo  {journal} {J. Phys. Soc. Jpn.}\ }\textbf {\bibinfo
  {volume} {85}},\ \bibinfo {pages} {112001} (\bibinfo {year}
  {2016})}\BibitemShut {NoStop}%
\bibitem [{\citenamefont {Nagaosa}\ and\ \citenamefont
  {Tokura}(2013)}]{nagaosa2013natnano}%
  \BibitemOpen
  \bibfield  {author} {\bibinfo {author} {\bibfnamefont {N.}~\bibnamefont
  {Nagaosa}}\ and\ \bibinfo {author} {\bibfnamefont {Y.}~\bibnamefont
  {Tokura}},\ }\bibfield  {title} {\bibinfo {title} {{Topological properties
  and dynamics of magnetic skyrmions}},\ }\href
  {https://doi.org/10.1038/nnano.2013.243} {\bibfield  {journal} {\bibinfo
  {journal} {Nat. Nano.}\ }\textbf {\bibinfo {volume} {8}},\ \bibinfo {pages}
  {899} (\bibinfo {year} {2013})}\BibitemShut {NoStop}%
\bibitem [{\citenamefont {Bednorz}\ and\ \citenamefont
  {M\"{u}ller}(1986)}]{bednorz1986zpb}%
  \BibitemOpen
  \bibfield  {author} {\bibinfo {author} {\bibfnamefont {J.~G.}\ \bibnamefont
  {Bednorz}}\ and\ \bibinfo {author} {\bibfnamefont {K.~A.}\ \bibnamefont
  {M\"{u}ller}},\ }\bibfield  {title} {\bibinfo {title} {{Possible high$T_{c}$
  superconductivity in the Ba-La-Cu-O system}},\ }\href
  {https://doi.org/10.1007/BF01303701} {\bibfield  {journal} {\bibinfo
  {journal} {Z. Phys. B}\ }\textbf {\bibinfo {volume} {64}},\ \bibinfo {pages}
  {189} (\bibinfo {year} {1986})}\BibitemShut {NoStop}%
\bibitem [{\citenamefont {Wu}\ \emph {et~al.}(2011)\citenamefont {Wu},
  \citenamefont {Mayaffre}, \citenamefont {Kr\"{a}mer}, \citenamefont
  {Horvati\'{c}}, \citenamefont {Berthier}, \citenamefont {Hardy},
  \citenamefont {Liang}, \citenamefont {Bonn},\ and\ \citenamefont
  {Julien}}]{wu2011nature}%
  \BibitemOpen
  \bibfield  {author} {\bibinfo {author} {\bibfnamefont {T.}~\bibnamefont
  {Wu}}, \bibinfo {author} {\bibfnamefont {H.}~\bibnamefont {Mayaffre}},
  \bibinfo {author} {\bibfnamefont {S.}~\bibnamefont {Kr\"{a}mer}}, \bibinfo
  {author} {\bibfnamefont {M.}~\bibnamefont {Horvati\'{c}}}, \bibinfo {author}
  {\bibfnamefont {C.}~\bibnamefont {Berthier}}, \bibinfo {author}
  {\bibfnamefont {W.~N.}\ \bibnamefont {Hardy}}, \bibinfo {author}
  {\bibfnamefont {R.}~\bibnamefont {Liang}}, \bibinfo {author} {\bibfnamefont
  {D.~A.}\ \bibnamefont {Bonn}},\ and\ \bibinfo {author} {\bibfnamefont
  {M.-H.}\ \bibnamefont {Julien}},\ }\bibfield  {title} {\bibinfo {title}
  {{Magnetic-field-induced charge-stripe order in the high-temperature
  superconductor YBa$_{2}$Cu$_{3}$O$_{y}$}},\ }\href
  {https://doi.org/10.1038/nature10345} {\bibfield  {journal} {\bibinfo
  {journal} {Nature}\ }\textbf {\bibinfo {volume} {477}},\ \bibinfo {pages}
  {191} (\bibinfo {year} {2011})}\BibitemShut {NoStop}%
\bibitem [{\citenamefont {Jin}\ \emph {et~al.}(1994)\citenamefont {Jin},
  \citenamefont {Tiefel}, \citenamefont {McCormack}, \citenamefont {Fastnacht},
  \citenamefont {Ramesh},\ and\ \citenamefont {Chen}}]{jin1994science}%
  \BibitemOpen
  \bibfield  {author} {\bibinfo {author} {\bibfnamefont {S.}~\bibnamefont
  {Jin}}, \bibinfo {author} {\bibfnamefont {T.~H.}\ \bibnamefont {Tiefel}},
  \bibinfo {author} {\bibfnamefont {M.}~\bibnamefont {McCormack}}, \bibinfo
  {author} {\bibfnamefont {R.~A.}\ \bibnamefont {Fastnacht}}, \bibinfo {author}
  {\bibfnamefont {R.}~\bibnamefont {Ramesh}},\ and\ \bibinfo {author}
  {\bibfnamefont {L.~H.}\ \bibnamefont {Chen}},\ }\bibfield  {title} {\bibinfo
  {title} {{Thousandfold Change in Resistivity in Magnetoresistive La-Ca-Mn-O
  Films}},\ }\href
  {https://www.science.org/doi/abs/10.1126/science.264.5157.413} {\bibfield
  {journal} {\bibinfo  {journal} {Science}\ }\textbf {\bibinfo {volume}
  {264}},\ \bibinfo {pages} {413} (\bibinfo {year} {1994})}\BibitemShut
  {NoStop}%
\bibitem [{\citenamefont {Uehara}\ \emph {et~al.}(1999)\citenamefont {Uehara},
  \citenamefont {Mori}, \citenamefont {Chen},\ and\ \citenamefont
  {Cheong}}]{uehara1999nature}%
  \BibitemOpen
  \bibfield  {author} {\bibinfo {author} {\bibfnamefont {M.}~\bibnamefont
  {Uehara}}, \bibinfo {author} {\bibfnamefont {S.}~\bibnamefont {Mori}},
  \bibinfo {author} {\bibfnamefont {C.~H.}\ \bibnamefont {Chen}},\ and\
  \bibinfo {author} {\bibfnamefont {S.-W.}\ \bibnamefont {Cheong}},\ }\bibfield
   {title} {\bibinfo {title} {{Percolative phase separation underlies colossal
  magnetoresistance in mixed-valent manganites}},\ }\href
  {https://doi.org/10.1038/21142} {\bibfield  {journal} {\bibinfo  {journal}
  {Nature}\ }\textbf {\bibinfo {volume} {399}},\ \bibinfo {pages} {560}
  (\bibinfo {year} {1999})}\BibitemShut {NoStop}%
\bibitem [{\citenamefont {Cao}\ \emph {et~al.}(2018)\citenamefont {Cao},
  \citenamefont {Fatemi}, \citenamefont {Fang}, \citenamefont {Watanabe},
  \citenamefont {Taniguchi}, \citenamefont {Kaxiras},\ and\ \citenamefont
  {Jarillo-Herrero}}]{cao2018nature}%
  \BibitemOpen
  \bibfield  {author} {\bibinfo {author} {\bibfnamefont {Y.}~\bibnamefont
  {Cao}}, \bibinfo {author} {\bibfnamefont {V.}~\bibnamefont {Fatemi}},
  \bibinfo {author} {\bibfnamefont {S.}~\bibnamefont {Fang}}, \bibinfo {author}
  {\bibfnamefont {K.}~\bibnamefont {Watanabe}}, \bibinfo {author}
  {\bibfnamefont {T.}~\bibnamefont {Taniguchi}}, \bibinfo {author}
  {\bibfnamefont {E.}~\bibnamefont {Kaxiras}},\ and\ \bibinfo {author}
  {\bibfnamefont {P.}~\bibnamefont {Jarillo-Herrero}},\ }\bibfield  {title}
  {\bibinfo {title} {{Unconventional superconductivity in magic-angle graphene
  superlattices}},\ }\href {https://doi.org/10.1038/nature26160} {\bibfield
  {journal} {\bibinfo  {journal} {Nature}\ }\textbf {\bibinfo {volume} {556}},\
  \bibinfo {pages} {43} (\bibinfo {year} {2018})}\BibitemShut {NoStop}%
\bibitem [{\citenamefont {Metzner}\ and\ \citenamefont
  {Vollhardt}(1989)}]{metzner1989prl}%
  \BibitemOpen
  \bibfield  {author} {\bibinfo {author} {\bibfnamefont {W.}~\bibnamefont
  {Metzner}}\ and\ \bibinfo {author} {\bibfnamefont {D.}~\bibnamefont
  {Vollhardt}},\ }\bibfield  {title} {\bibinfo {title} {{Correlated Lattice
  Fermions in $d=\ensuremath{\infty}$ Dimensions}},\ }\href
  {https://link.aps.org/doi/10.1103/PhysRevLett.62.324} {\bibfield  {journal}
  {\bibinfo  {journal} {Phys. Rev. Lett.}\ }\textbf {\bibinfo {volume} {62}},\
  \bibinfo {pages} {324} (\bibinfo {year} {1989})}\BibitemShut {NoStop}%
\bibitem [{\citenamefont {Georges}\ and\ \citenamefont
  {Kotliar}(1992)}]{georges1992prb}%
  \BibitemOpen
  \bibfield  {author} {\bibinfo {author} {\bibfnamefont {A.}~\bibnamefont
  {Georges}}\ and\ \bibinfo {author} {\bibfnamefont {G.}~\bibnamefont
  {Kotliar}},\ }\bibfield  {title} {\bibinfo {title} {{Hubbard model in
  infinite dimensions}},\ }\href
  {https://link.aps.org/doi/10.1103/PhysRevB.45.6479} {\bibfield  {journal}
  {\bibinfo  {journal} {Phys. Rev. B}\ }\textbf {\bibinfo {volume} {45}},\
  \bibinfo {pages} {6479} (\bibinfo {year} {1992})}\BibitemShut {NoStop}%
\bibitem [{\citenamefont {Lichtenstein}\ and\ \citenamefont
  {Katsnelson}(2000)}]{lichtenstein2000prb}%
  \BibitemOpen
  \bibfield  {author} {\bibinfo {author} {\bibfnamefont {A.~I.}\ \bibnamefont
  {Lichtenstein}}\ and\ \bibinfo {author} {\bibfnamefont {M.~I.}\ \bibnamefont
  {Katsnelson}},\ }\bibfield  {title} {\bibinfo {title} {{Antiferromagnetism
  and d-wave superconductivity in cuprates: A cluster dynamical mean-field
  theory}},\ }\href {https://link.aps.org/doi/10.1103/PhysRevB.62.R9283}
  {\bibfield  {journal} {\bibinfo  {journal} {Phys. Rev. B}\ }\textbf {\bibinfo
  {volume} {62}},\ \bibinfo {pages} {R9283} (\bibinfo {year}
  {2000})}\BibitemShut {NoStop}%
\bibitem [{\citenamefont {Kotliar}\ \emph {et~al.}(2001)\citenamefont
  {Kotliar}, \citenamefont {Savrasov}, \citenamefont {P\'alsson},\ and\
  \citenamefont {Biroli}}]{kotliar2001prl}%
  \BibitemOpen
  \bibfield  {author} {\bibinfo {author} {\bibfnamefont {G.}~\bibnamefont
  {Kotliar}}, \bibinfo {author} {\bibfnamefont {S.~Y.}\ \bibnamefont
  {Savrasov}}, \bibinfo {author} {\bibfnamefont {G.}~\bibnamefont
  {P\'alsson}},\ and\ \bibinfo {author} {\bibfnamefont {G.}~\bibnamefont
  {Biroli}},\ }\bibfield  {title} {\bibinfo {title} {{Cellular Dynamical Mean
  Field Approach to Strongly Correlated Systems}},\ }\href
  {https://link.aps.org/doi/10.1103/PhysRevLett.87.186401} {\bibfield
  {journal} {\bibinfo  {journal} {Phys. Rev. Lett.}\ }\textbf {\bibinfo
  {volume} {87}},\ \bibinfo {pages} {186401} (\bibinfo {year}
  {2001})}\BibitemShut {NoStop}%
\bibitem [{\citenamefont {S\'en\'echal}\ \emph {et~al.}(2000)\citenamefont
  {S\'en\'echal}, \citenamefont {Perez},\ and\ \citenamefont
  {Pioro-Ladri\`ere}}]{senechal2000prl}%
  \BibitemOpen
  \bibfield  {author} {\bibinfo {author} {\bibfnamefont {D.}~\bibnamefont
  {S\'en\'echal}}, \bibinfo {author} {\bibfnamefont {D.}~\bibnamefont
  {Perez}},\ and\ \bibinfo {author} {\bibfnamefont {M.}~\bibnamefont
  {Pioro-Ladri\`ere}},\ }\bibfield  {title} {\bibinfo {title} {{Spectral Weight
  of the Hubbard Model through Cluster Perturbation Theory}},\ }\href
  {https://link.aps.org/doi/10.1103/PhysRevLett.84.522} {\bibfield  {journal}
  {\bibinfo  {journal} {Phys. Rev. Lett.}\ }\textbf {\bibinfo {volume} {84}},\
  \bibinfo {pages} {522} (\bibinfo {year} {2000})}\BibitemShut {NoStop}%
\bibitem [{\citenamefont {S\'en\'echal}\ \emph {et~al.}(2002)\citenamefont
  {S\'en\'echal}, \citenamefont {Perez},\ and\ \citenamefont
  {Plouffe}}]{senechal2002prb}%
  \BibitemOpen
  \bibfield  {author} {\bibinfo {author} {\bibfnamefont {D.}~\bibnamefont
  {S\'en\'echal}}, \bibinfo {author} {\bibfnamefont {D.}~\bibnamefont
  {Perez}},\ and\ \bibinfo {author} {\bibfnamefont {D.}~\bibnamefont
  {Plouffe}},\ }\bibfield  {title} {\bibinfo {title} {{Cluster perturbation
  theory for Hubbard models}},\ }\href
  {https://link.aps.org/doi/10.1103/PhysRevB.66.075129} {\bibfield  {journal}
  {\bibinfo  {journal} {Phys. Rev. B}\ }\textbf {\bibinfo {volume} {66}},\
  \bibinfo {pages} {075129} (\bibinfo {year} {2002})}\BibitemShut {NoStop}%
\bibitem [{\citenamefont {White}(1992)}]{white1992prl}%
  \BibitemOpen
  \bibfield  {author} {\bibinfo {author} {\bibfnamefont {S.~R.}\ \bibnamefont
  {White}},\ }\bibfield  {title} {\bibinfo {title} {{Density matrix formulation
  for quantum renormalization groups}},\ }\href
  {https://link.aps.org/doi/10.1103/PhysRevLett.69.2863} {\bibfield  {journal}
  {\bibinfo  {journal} {Phys. Rev. Lett.}\ }\textbf {\bibinfo {volume} {69}},\
  \bibinfo {pages} {2863} (\bibinfo {year} {1992})}\BibitemShut {NoStop}%
\bibitem [{\citenamefont {White}(1993)}]{white1993prb}%
  \BibitemOpen
  \bibfield  {author} {\bibinfo {author} {\bibfnamefont {S.~R.}\ \bibnamefont
  {White}},\ }\bibfield  {title} {\bibinfo {title} {{Density-matrix algorithms
  for quantum renormalization groups}},\ }\href
  {https://link.aps.org/doi/10.1103/PhysRevB.48.10345} {\bibfield  {journal}
  {\bibinfo  {journal} {Phys. Rev. B}\ }\textbf {\bibinfo {volume} {48}},\
  \bibinfo {pages} {10345} (\bibinfo {year} {1993})}\BibitemShut {NoStop}%
\bibitem [{\citenamefont {Schollw\"ock}(2011)}]{schollwock2011ap}%
  \BibitemOpen
  \bibfield  {author} {\bibinfo {author} {\bibfnamefont {U.}~\bibnamefont
  {Schollw\"ock}},\ }\bibfield  {title} {\bibinfo {title} {{The density-matrix
  renormalization group in the age of matrix product states}},\ }\href
  {http://www.sciencedirect.com/science/article/pii/S0003491610001752}
  {\bibfield  {journal} {\bibinfo  {journal} {Ann. Phys.}\ }\textbf {\bibinfo
  {volume} {326}},\ \bibinfo {pages} {96} (\bibinfo {year} {2011})}\BibitemShut
  {NoStop}%
\bibitem [{\citenamefont {Nishio}\ \emph {et~al.}(2004)\citenamefont {Nishio},
  \citenamefont {Maeshima}, \citenamefont {Gendiar},\ and\ \citenamefont
  {Nishino}}]{nishio2004arxiv}%
  \BibitemOpen
  \bibfield  {author} {\bibinfo {author} {\bibfnamefont {Y.}~\bibnamefont
  {Nishio}}, \bibinfo {author} {\bibfnamefont {N.}~\bibnamefont {Maeshima}},
  \bibinfo {author} {\bibfnamefont {A.}~\bibnamefont {Gendiar}},\ and\ \bibinfo
  {author} {\bibfnamefont {T.}~\bibnamefont {Nishino}},\ }\bibfield  {title}
  {\bibinfo {title} {{Tensor Product Variational Formulation for Quantum
  Systems}},\ }\href {https://arxiv.org/abs/cond-mat/0401115} {\bibfield
  {journal} {\bibinfo  {journal} {arXiv:cond-mat/0401115}\ } (\bibinfo {year}
  {2004})}\BibitemShut {NoStop}%
\bibitem [{\citenamefont {Jordan}\ \emph {et~al.}(2008)\citenamefont {Jordan},
  \citenamefont {Or\'us}, \citenamefont {Vidal}, \citenamefont {Verstraete},\
  and\ \citenamefont {Cirac}}]{jordan2008prl}%
  \BibitemOpen
  \bibfield  {author} {\bibinfo {author} {\bibfnamefont {J.}~\bibnamefont
  {Jordan}}, \bibinfo {author} {\bibfnamefont {R.}~\bibnamefont {Or\'us}},
  \bibinfo {author} {\bibfnamefont {G.}~\bibnamefont {Vidal}}, \bibinfo
  {author} {\bibfnamefont {F.}~\bibnamefont {Verstraete}},\ and\ \bibinfo
  {author} {\bibfnamefont {J.~I.}\ \bibnamefont {Cirac}},\ }\bibfield  {title}
  {\bibinfo {title} {{Classical Simulation of Infinite-Size Quantum Lattice
  Systems in Two Spatial Dimensions}},\ }\href
  {https://link.aps.org/doi/10.1103/PhysRevLett.101.250602} {\bibfield
  {journal} {\bibinfo  {journal} {Phys. Rev. Lett.}\ }\textbf {\bibinfo
  {volume} {101}},\ \bibinfo {pages} {250602} (\bibinfo {year}
  {2008})}\BibitemShut {NoStop}%
\bibitem [{\citenamefont {Or\'us}(2014)}]{orus2014ap}%
  \BibitemOpen
  \bibfield  {author} {\bibinfo {author} {\bibfnamefont {R.}~\bibnamefont
  {Or\'us}},\ }\bibfield  {title} {\bibinfo {title} {{A practical introduction
  to tensor networks: Matrix product states and projected entangled pair
  states}},\ }\href
  {http://www.sciencedirect.com/science/article/pii/S0003491614001596}
  {\bibfield  {journal} {\bibinfo  {journal} {Ann. Phys.}\ }\textbf {\bibinfo
  {volume} {349}},\ \bibinfo {pages} {117} (\bibinfo {year}
  {2014})}\BibitemShut {NoStop}%
\bibitem [{\citenamefont {Shibata}\ and\ \citenamefont
  {Hotta}(2011)}]{shibata2011prb}%
  \BibitemOpen
  \bibfield  {author} {\bibinfo {author} {\bibfnamefont {N.}~\bibnamefont
  {Shibata}}\ and\ \bibinfo {author} {\bibfnamefont {C.}~\bibnamefont
  {Hotta}},\ }\bibfield  {title} {\bibinfo {title} {{Boundary effects in the
  density-matrix renormalization group calculation}},\ }\href
  {https://link.aps.org/doi/10.1103/PhysRevB.84.115116} {\bibfield  {journal}
  {\bibinfo  {journal} {Phys. Rev. B}\ }\textbf {\bibinfo {volume} {84}},\
  \bibinfo {pages} {115116} (\bibinfo {year} {2011})}\BibitemShut {NoStop}%
\bibitem [{\citenamefont {Hotta}\ and\ \citenamefont
  {Shibata}(2012)}]{hotta2012prb}%
  \BibitemOpen
  \bibfield  {author} {\bibinfo {author} {\bibfnamefont {C.}~\bibnamefont
  {Hotta}}\ and\ \bibinfo {author} {\bibfnamefont {N.}~\bibnamefont
  {Shibata}},\ }\bibfield  {title} {\bibinfo {title} {{Grand canonical
  finite-size numerical approaches: A route to measuring bulk properties in an
  applied field}},\ }\href
  {https://link.aps.org/doi/10.1103/PhysRevB.86.041108} {\bibfield  {journal}
  {\bibinfo  {journal} {Phys. Rev. B}\ }\textbf {\bibinfo {volume} {86}},\
  \bibinfo {pages} {041108(R)} (\bibinfo {year} {2012})}\BibitemShut {NoStop}%
\bibitem [{\citenamefont {Hotta}\ \emph {et~al.}(2013)\citenamefont {Hotta},
  \citenamefont {Nishimoto},\ and\ \citenamefont {Shibata}}]{hotta2013prb}%
  \BibitemOpen
  \bibfield  {author} {\bibinfo {author} {\bibfnamefont {C.}~\bibnamefont
  {Hotta}}, \bibinfo {author} {\bibfnamefont {S.}~\bibnamefont {Nishimoto}},\
  and\ \bibinfo {author} {\bibfnamefont {N.}~\bibnamefont {Shibata}},\
  }\bibfield  {title} {\bibinfo {title} {{Grand canonical finite size numerical
  approaches in one and two dimensions: Real space energy renormalization and
  edge state generation}},\ }\href
  {https://link.aps.org/doi/10.1103/PhysRevB.87.115128} {\bibfield  {journal}
  {\bibinfo  {journal} {Phys. Rev. B}\ }\textbf {\bibinfo {volume} {87}},\
  \bibinfo {pages} {115128} (\bibinfo {year} {2013})}\BibitemShut {NoStop}%
\bibitem [{\citenamefont {Gros}\ \emph {et~al.}(1987)\citenamefont {Gros},
  \citenamefont {Joynt},\ and\ \citenamefont {Rice}}]{gros1987prb}%
  \BibitemOpen
  \bibfield  {author} {\bibinfo {author} {\bibfnamefont {C.}~\bibnamefont
  {Gros}}, \bibinfo {author} {\bibfnamefont {R.}~\bibnamefont {Joynt}},\ and\
  \bibinfo {author} {\bibfnamefont {T.~M.}\ \bibnamefont {Rice}},\ }\bibfield
  {title} {\bibinfo {title} {{Antiferromagnetic correlations in
  almost-localized Fermi liquids}},\ }\href
  {https://link.aps.org/doi/10.1103/PhysRevB.36.381} {\bibfield  {journal}
  {\bibinfo  {journal} {Phys. Rev. B}\ }\textbf {\bibinfo {volume} {36}},\
  \bibinfo {pages} {381} (\bibinfo {year} {1987})}\BibitemShut {NoStop}%
\bibitem [{\citenamefont {Gros}(1988)}]{gros1988prb}%
  \BibitemOpen
  \bibfield  {author} {\bibinfo {author} {\bibfnamefont {C.}~\bibnamefont
  {Gros}},\ }\bibfield  {title} {\bibinfo {title} {{Superconductivity in
  correlated wave functions}},\ }\href
  {https://link.aps.org/doi/10.1103/PhysRevB.38.931} {\bibfield  {journal}
  {\bibinfo  {journal} {Phys. Rev. B}\ }\textbf {\bibinfo {volume} {38}},\
  \bibinfo {pages} {931} (\bibinfo {year} {1988})}\BibitemShut {NoStop}%
\bibitem [{\citenamefont {Giamarchi}\ and\ \citenamefont
  {Lhuillier}(1991)}]{giamarchi1991prb}%
  \BibitemOpen
  \bibfield  {author} {\bibinfo {author} {\bibfnamefont {T.}~\bibnamefont
  {Giamarchi}}\ and\ \bibinfo {author} {\bibfnamefont {C.}~\bibnamefont
  {Lhuillier}},\ }\bibfield  {title} {\bibinfo {title} {{Phase diagrams of the
  two-dimensional Hubbard and t-J models by a variational Monte Carlo
  method}},\ }\href {https://link.aps.org/doi/10.1103/PhysRevB.43.12943}
  {\bibfield  {journal} {\bibinfo  {journal} {Phys. Rev. B}\ }\textbf {\bibinfo
  {volume} {43}},\ \bibinfo {pages} {12943} (\bibinfo {year}
  {1991})}\BibitemShut {NoStop}%
\bibitem [{\citenamefont {Dev}\ and\ \citenamefont {Jain}(1992)}]{dev1992prb}%
  \BibitemOpen
  \bibfield  {author} {\bibinfo {author} {\bibfnamefont {G.}~\bibnamefont
  {Dev}}\ and\ \bibinfo {author} {\bibfnamefont {J.~K.}\ \bibnamefont {Jain}},\
  }\bibfield  {title} {\bibinfo {title} {{Jastrow-Slater trial wave functions
  for the fractional quantum Hall effect: Results for few-particle systems}},\
  }\href {https://link.aps.org/doi/10.1103/PhysRevB.45.1223} {\bibfield
  {journal} {\bibinfo  {journal} {Phys. Rev. B}\ }\textbf {\bibinfo {volume}
  {45}},\ \bibinfo {pages} {1223} (\bibinfo {year} {1992})}\BibitemShut
  {NoStop}%
\bibitem [{\citenamefont {Sorella}(2005)}]{sorella2005prb}%
  \BibitemOpen
  \bibfield  {author} {\bibinfo {author} {\bibfnamefont {S.}~\bibnamefont
  {Sorella}},\ }\bibfield  {title} {\bibinfo {title} {{Wave function
  optimization in the variational Monte Carlo method}},\ }\href
  {https://link.aps.org/doi/10.1103/PhysRevB.71.241103} {\bibfield  {journal}
  {\bibinfo  {journal} {Phys. Rev. B}\ }\textbf {\bibinfo {volume} {71}},\
  \bibinfo {pages} {241103(R)} (\bibinfo {year} {2005})}\BibitemShut {NoStop}%
\bibitem [{\citenamefont {Suzuki}(1976)}]{suzuki1976ptp}%
  \BibitemOpen
  \bibfield  {author} {\bibinfo {author} {\bibfnamefont {M.}~\bibnamefont
  {Suzuki}},\ }\bibfield  {title} {\bibinfo {title} {{Relationship between
  d-Dimensional Quantal Spin Systems and (d+1)-Dimensional Ising Systems:
  Equivalence, Critical Exponents and Systematic Approximants of the Partition
  Function and Spin Correlations}},\ }\href
  {https://doi.org/10.1143/PTP.56.1454} {\bibfield  {journal} {\bibinfo
  {journal} {Prog. Theor. Phys.}\ }\textbf {\bibinfo {volume} {56}},\ \bibinfo
  {pages} {1454} (\bibinfo {year} {1976})}\BibitemShut {NoStop}%
\bibitem [{\citenamefont {Foulkes}\ \emph {et~al.}(2001)\citenamefont
  {Foulkes}, \citenamefont {Mitas}, \citenamefont {Needs},\ and\ \citenamefont
  {Rajagopal}}]{foulkes2001rmp}%
  \BibitemOpen
  \bibfield  {author} {\bibinfo {author} {\bibfnamefont {W.~M.~C.}\
  \bibnamefont {Foulkes}}, \bibinfo {author} {\bibfnamefont {L.}~\bibnamefont
  {Mitas}}, \bibinfo {author} {\bibfnamefont {R.~J.}\ \bibnamefont {Needs}},\
  and\ \bibinfo {author} {\bibfnamefont {G.}~\bibnamefont {Rajagopal}},\
  }\bibfield  {title} {\bibinfo {title} {{Quantum Monte Carlo simulations of
  solids}},\ }\href {https://link.aps.org/doi/10.1103/RevModPhys.73.33}
  {\bibfield  {journal} {\bibinfo  {journal} {Rev. Mod. Phys.}\ }\textbf
  {\bibinfo {volume} {73}},\ \bibinfo {pages} {33} (\bibinfo {year}
  {2001})}\BibitemShut {NoStop}%
\bibitem [{\citenamefont {Okubo}\ \emph {et~al.}(2012)\citenamefont {Okubo},
  \citenamefont {Chung},\ and\ \citenamefont {Kawamura}}]{okubo2021prl}%
  \BibitemOpen
  \bibfield  {author} {\bibinfo {author} {\bibfnamefont {T.}~\bibnamefont
  {Okubo}}, \bibinfo {author} {\bibfnamefont {S.}~\bibnamefont {Chung}},\ and\
  \bibinfo {author} {\bibfnamefont {H.}~\bibnamefont {Kawamura}},\ }\bibfield
  {title} {\bibinfo {title} {{Multiple-$q$ States and the Skyrmion Lattice of
  the Triangular-Lattice Heisenberg Antiferromagnet under Magnetic Fields}},\
  }\href {https://link.aps.org/doi/10.1103/PhysRevLett.108.017206} {\bibfield
  {journal} {\bibinfo  {journal} {Phys. Rev. Lett.}\ }\textbf {\bibinfo
  {volume} {108}},\ \bibinfo {pages} {017206} (\bibinfo {year}
  {2012})}\BibitemShut {NoStop}%
\bibitem [{\citenamefont {Gendiar}\ \emph {et~al.}(2009)\citenamefont
  {Gendiar}, \citenamefont {Krcmar},\ and\ \citenamefont
  {Nishino}}]{gendiar2009ptp}%
  \BibitemOpen
  \bibfield  {author} {\bibinfo {author} {\bibfnamefont {A.}~\bibnamefont
  {Gendiar}}, \bibinfo {author} {\bibfnamefont {R.}~\bibnamefont {Krcmar}},\
  and\ \bibinfo {author} {\bibfnamefont {T.}~\bibnamefont {Nishino}},\
  }\bibfield  {title} {\bibinfo {title} {{Spherical Deformation for
  One-Dimensional Quantum Systems}},\ }\href
  {https://doi.org/10.1143/PTP.122.953} {\bibfield  {journal} {\bibinfo
  {journal} {Prog. Theor. Phys.}\ }\textbf {\bibinfo {volume} {122}},\ \bibinfo
  {pages} {953} (\bibinfo {year} {2009})}\BibitemShut {NoStop}%
\bibitem [{\citenamefont {Hikihara}\ and\ \citenamefont
  {Nishino}(2011)}]{hikihara2011prb}%
  \BibitemOpen
  \bibfield  {author} {\bibinfo {author} {\bibfnamefont {T.}~\bibnamefont
  {Hikihara}}\ and\ \bibinfo {author} {\bibfnamefont {T.}~\bibnamefont
  {Nishino}},\ }\bibfield  {title} {\bibinfo {title} {{Connecting distant ends
  of one-dimensional critical systems by a sine-square deformation}},\ }\href
  {https://link.aps.org/doi/10.1103/PhysRevB.83.060414} {\bibfield  {journal}
  {\bibinfo  {journal} {Phys. Rev. B}\ }\textbf {\bibinfo {volume} {83}},\
  \bibinfo {pages} {060414(R)} (\bibinfo {year} {2011})}\BibitemShut {NoStop}%
\bibitem [{\citenamefont {Gendiar}\ \emph {et~al.}(2011)\citenamefont
  {Gendiar}, \citenamefont {Dani\ifmmode~\check{s}\else \v{s}\fi{}ka},
  \citenamefont {Lee},\ and\ \citenamefont {Nishino}}]{gendiar2011pra}%
  \BibitemOpen
  \bibfield  {author} {\bibinfo {author} {\bibfnamefont {A.}~\bibnamefont
  {Gendiar}}, \bibinfo {author} {\bibfnamefont {M.}~\bibnamefont
  {Dani\ifmmode~\check{s}\else \v{s}\fi{}ka}}, \bibinfo {author} {\bibfnamefont
  {Y.}~\bibnamefont {Lee}},\ and\ \bibinfo {author} {\bibfnamefont
  {T.}~\bibnamefont {Nishino}},\ }\bibfield  {title} {\bibinfo {title}
  {{Suppression of finite-size effects in one-dimensional correlated
  systems}},\ }\href {https://link.aps.org/doi/10.1103/PhysRevA.83.052118}
  {\bibfield  {journal} {\bibinfo  {journal} {Phys. Rev. A}\ }\textbf {\bibinfo
  {volume} {83}},\ \bibinfo {pages} {052118} (\bibinfo {year}
  {2011})}\BibitemShut {NoStop}%
\bibitem [{\citenamefont {Tada}(2015)}]{tada2015mpl}%
  \BibitemOpen
  \bibfield  {author} {\bibinfo {author} {\bibfnamefont {T.}~\bibnamefont
  {Tada}},\ }\bibfield  {title} {\bibinfo {title} {{Sine-square deformation and
  its relevance to string theory}},\ }\href
  {https://doi.org/10.1142/s0217732315500923} {\bibfield  {journal} {\bibinfo
  {journal} {Mod. Phys. Lett. A}\ }\textbf {\bibinfo {volume} {30}},\ \bibinfo
  {pages} {1550092} (\bibinfo {year} {2015})}\BibitemShut {NoStop}%
\bibitem [{\citenamefont {Okunishi}\ and\ \citenamefont
  {Katsura}(2015)}]{okunishi2015jpa}%
  \BibitemOpen
  \bibfield  {author} {\bibinfo {author} {\bibfnamefont {K.}~\bibnamefont
  {Okunishi}}\ and\ \bibinfo {author} {\bibfnamefont {H.}~\bibnamefont
  {Katsura}},\ }\bibfield  {title} {\bibinfo {title} {{Sine-square deformation
  and supersymmetric quantum mechanics}},\ }\href
  {https://doi.org/10.1088/1751-8113/48/44/445208} {\bibfield  {journal}
  {\bibinfo  {journal} {J. Phys. A: Math. Theor.}\ }\textbf {\bibinfo {volume}
  {48}},\ \bibinfo {pages} {445208} (\bibinfo {year} {2015})}\BibitemShut
  {NoStop}%
\bibitem [{\citenamefont {Yonaga}\ and\ \citenamefont
  {Shibata}(2015)}]{yonaga2015jpsj}%
  \BibitemOpen
  \bibfield  {author} {\bibinfo {author} {\bibfnamefont {K.}~\bibnamefont
  {Yonaga}}\ and\ \bibinfo {author} {\bibfnamefont {N.}~\bibnamefont
  {Shibata}},\ }\bibfield  {title} {\bibinfo {title} {{Ground State Phase
  Diagram of Twisted Three-Leg Spin Tube in Magnetic Field}},\ }\href
  {https://doi.org/10.7566/JPSJ.84.094706} {\bibfield  {journal} {\bibinfo
  {journal} {J. Phys. Soc. Jpn.}\ }\textbf {\bibinfo {volume} {84}},\ \bibinfo
  {pages} {094706} (\bibinfo {year} {2015})}\BibitemShut {NoStop}%
\bibitem [{\citenamefont {Okunishi}(2016)}]{okumishi2016ptep}%
  \BibitemOpen
  \bibfield  {author} {\bibinfo {author} {\bibfnamefont {K.}~\bibnamefont
  {Okunishi}},\ }\bibfield  {title} {\bibinfo {title} {{Sine-square deformation
  and Möbius quantization of 2D conformal field theory}},\ }\href
  {https://doi.org/10.1093/ptep/ptw060} {\bibfield  {journal} {\bibinfo
  {journal} {Prog. Theor. Exp. Phys.}\ }\textbf {\bibinfo {volume} {2016}},\
  \bibinfo {pages} {2050} (\bibinfo {year} {2016})}\BibitemShut {NoStop}%
\bibitem [{\citenamefont {Wen}\ \emph {et~al.}(2016)\citenamefont {Wen},
  \citenamefont {Ryu},\ and\ \citenamefont {Ludwig}}]{wen2016prb}%
  \BibitemOpen
  \bibfield  {author} {\bibinfo {author} {\bibfnamefont {X.}~\bibnamefont
  {Wen}}, \bibinfo {author} {\bibfnamefont {S.}~\bibnamefont {Ryu}},\ and\
  \bibinfo {author} {\bibfnamefont {A.~W.~W.}\ \bibnamefont {Ludwig}},\
  }\bibfield  {title} {\bibinfo {title} {{Evolution operators in conformal
  field theories and conformal mappings: Entanglement Hamiltonian, the
  sine-square deformation, and others}},\ }\href
  {https://link.aps.org/doi/10.1103/PhysRevB.93.235119} {\bibfield  {journal}
  {\bibinfo  {journal} {Phys. Rev. B}\ }\textbf {\bibinfo {volume} {93}},\
  \bibinfo {pages} {235119} (\bibinfo {year} {2016})}\BibitemShut {NoStop}%
\bibitem [{\citenamefont {Morita}\ and\ \citenamefont
  {Shibata}(2016{\natexlab{a}})}]{morita2016prb}%
  \BibitemOpen
  \bibfield  {author} {\bibinfo {author} {\bibfnamefont {K.}~\bibnamefont
  {Morita}}\ and\ \bibinfo {author} {\bibfnamefont {N.}~\bibnamefont
  {Shibata}},\ }\bibfield  {title} {\bibinfo {title} {{Multiple magnetization
  plateaus and magnetic structures in the $S=\frac{1}{2}$ Heisenberg model on
  the checkerboard lattice}},\ }\href
  {https://link.aps.org/doi/10.1103/PhysRevB.94.140404} {\bibfield  {journal}
  {\bibinfo  {journal} {Phys. Rev. B}\ }\textbf {\bibinfo {volume} {94}},\
  \bibinfo {pages} {140404(R)} (\bibinfo {year}
  {2016}{\natexlab{a}})}\BibitemShut {NoStop}%
\bibitem [{\citenamefont {Morita}\ and\ \citenamefont
  {Shibata}(2016{\natexlab{b}})}]{morita2016jpsj}%
  \BibitemOpen
  \bibfield  {author} {\bibinfo {author} {\bibfnamefont {K.}~\bibnamefont
  {Morita}}\ and\ \bibinfo {author} {\bibfnamefont {N.}~\bibnamefont
  {Shibata}},\ }\bibfield  {title} {\bibinfo {title} {{Field-Induced Quantum
  Phase Transitions in S = 1/2 J1–J2 Heisenberg Model on Square Lattice}},\
  }\href {https://doi.org/10.7566/JPSJ.85.094708} {\bibfield  {journal}
  {\bibinfo  {journal} {J. Phys. Soc. Jpn.}\ }\textbf {\bibinfo {volume}
  {85}},\ \bibinfo {pages} {094708} (\bibinfo {year}
  {2016}{\natexlab{b}})}\BibitemShut {NoStop}%
\bibitem [{\citenamefont {Hotta}\ and\ \citenamefont
  {Asano}(2018)}]{hotta2018prb}%
  \BibitemOpen
  \bibfield  {author} {\bibinfo {author} {\bibfnamefont {C.}~\bibnamefont
  {Hotta}}\ and\ \bibinfo {author} {\bibfnamefont {K.}~\bibnamefont {Asano}},\
  }\bibfield  {title} {\bibinfo {title} {{Magnetic susceptibility of quantum
  spin systems calculated by sine square deformation: One-dimensional, square
  lattice, and kagome lattice Heisenberg antiferromagnets}},\ }\href
  {https://link.aps.org/doi/10.1103/PhysRevB.98.140405} {\bibfield  {journal}
  {\bibinfo  {journal} {Phys. Rev. B}\ }\textbf {\bibinfo {volume} {98}},\
  \bibinfo {pages} {140405(R)} (\bibinfo {year} {2018})}\BibitemShut {NoStop}%
\bibitem [{\citenamefont {Kadosawa}\ \emph {et~al.}(2020)\citenamefont
  {Kadosawa}, \citenamefont {Nishimoto}, \citenamefont {Sugimoto},\ and\
  \citenamefont {Ohta}}]{kadosawa2020jpsj}%
  \BibitemOpen
  \bibfield  {author} {\bibinfo {author} {\bibfnamefont {M.}~\bibnamefont
  {Kadosawa}}, \bibinfo {author} {\bibfnamefont {S.}~\bibnamefont {Nishimoto}},
  \bibinfo {author} {\bibfnamefont {K.}~\bibnamefont {Sugimoto}},\ and\
  \bibinfo {author} {\bibfnamefont {Y.}~\bibnamefont {Ohta}},\ }\bibfield
  {title} {\bibinfo {title} {{Finite-Temperature Properties of Excitonic
  Condensation in the Extended Falicov–Kimball Model: Cluster
  Mean-Field-Theory Approach}},\ }\href
  {https://doi.org/10.7566/JPSJ.89.053706} {\bibfield  {journal} {\bibinfo
  {journal} {J. Phys. Soc. Jpn.}\ }\textbf {\bibinfo {volume} {89}},\ \bibinfo
  {pages} {053706} (\bibinfo {year} {2020})}\BibitemShut {NoStop}%
\bibitem [{\citenamefont {Hotta}\ \emph {et~al.}(2021)\citenamefont {Hotta},
  \citenamefont {Nakamaniwa},\ and\ \citenamefont {Nakamura}}]{hotta2021arxiv}%
  \BibitemOpen
  \bibfield  {author} {\bibinfo {author} {\bibfnamefont {C.}~\bibnamefont
  {Hotta}}, \bibinfo {author} {\bibfnamefont {T.}~\bibnamefont {Nakamaniwa}},\
  and\ \bibinfo {author} {\bibfnamefont {T.}~\bibnamefont {Nakamura}},\
  }\bibfield  {title} {\bibinfo {title} {Sine-square deformation applied to
  classical ising models},\ }\href
  {https://doi.org/10.1103/PhysRevE.104.034133} {\bibfield  {journal} {\bibinfo
   {journal} {Phys. Rev. E}\ }\textbf {\bibinfo {volume} {104}},\ \bibinfo
  {pages} {034133} (\bibinfo {year} {2021})}\BibitemShut {NoStop}%
\bibitem [{\citenamefont {Eguchi}\ \emph {et~al.}(2021)\citenamefont {Eguchi},
  \citenamefont {Oga}, \citenamefont {Katsura}, \citenamefont {Gendiar},\ and\
  \citenamefont {Nishino}}]{takuma2021arxiv}%
  \BibitemOpen
  \bibfield  {author} {\bibinfo {author} {\bibfnamefont {T.}~\bibnamefont
  {Eguchi}}, \bibinfo {author} {\bibfnamefont {S.}~\bibnamefont {Oga}},
  \bibinfo {author} {\bibfnamefont {H.}~\bibnamefont {Katsura}}, \bibinfo
  {author} {\bibfnamefont {A.}~\bibnamefont {Gendiar}},\ and\ \bibinfo {author}
  {\bibfnamefont {T.}~\bibnamefont {Nishino}},\ }\bibfield  {title} {\bibinfo
  {title} {{Energy Scale Deformation on Regular Polyhedra}},\ }\href
  {https://arxiv.org/abs/2109.10565} {\bibfield  {journal} {\bibinfo  {journal}
  {arXiv:2109.10565}\ } (\bibinfo {year} {2021})}\BibitemShut {NoStop}%
\bibitem [{\citenamefont {Katsura}(2011)}]{katsura2011jpa}%
  \BibitemOpen
  \bibfield  {author} {\bibinfo {author} {\bibfnamefont {H.}~\bibnamefont
  {Katsura}},\ }\bibfield  {title} {\bibinfo {title} {{Exact ground state of
  the sine-square deformed {XY} spin chain}},\ }\href
  {https://doi.org/10.1088/1751-8113/44/25/252001} {\bibfield  {journal}
  {\bibinfo  {journal} {J. Phys. A: Math. Theor.}\ }\textbf {\bibinfo {volume}
  {44}},\ \bibinfo {pages} {252001} (\bibinfo {year} {2011})}\BibitemShut
  {NoStop}%
\bibitem [{\citenamefont {Maruyama}\ \emph {et~al.}(2011)\citenamefont
  {Maruyama}, \citenamefont {Katsura},\ and\ \citenamefont
  {Hikihara}}]{maruyama2011prb}%
  \BibitemOpen
  \bibfield  {author} {\bibinfo {author} {\bibfnamefont {I.}~\bibnamefont
  {Maruyama}}, \bibinfo {author} {\bibfnamefont {H.}~\bibnamefont {Katsura}},\
  and\ \bibinfo {author} {\bibfnamefont {T.}~\bibnamefont {Hikihara}},\
  }\bibfield  {title} {\bibinfo {title} {{Sine-square deformation of free
  fermion systems in one and higher dimensions}},\ }\href
  {https://link.aps.org/doi/10.1103/PhysRevB.84.165132} {\bibfield  {journal}
  {\bibinfo  {journal} {Phys. Rev. B}\ }\textbf {\bibinfo {volume} {84}},\
  \bibinfo {pages} {165132} (\bibinfo {year} {2011})}\BibitemShut {NoStop}%
\bibitem [{\citenamefont {Katsura}(2012)}]{katsura2012jpa}%
  \BibitemOpen
  \bibfield  {author} {\bibinfo {author} {\bibfnamefont {H.}~\bibnamefont
  {Katsura}},\ }\bibfield  {title} {\bibinfo {title} {{Sine-square deformation
  of solvable spin chains and conformal field theories}},\ }\href
  {https://doi.org/10.1088/1751-8113/45/11/115003} {\bibfield  {journal}
  {\bibinfo  {journal} {J. Phys. A: Math. Theor.}\ }\textbf {\bibinfo {volume}
  {45}},\ \bibinfo {pages} {115003} (\bibinfo {year} {2012})}\BibitemShut
  {NoStop}%
\bibitem [{\citenamefont {Hikihara}\ and\ \citenamefont
  {Suzuki}(2013)}]{hikihara2013pra}%
  \BibitemOpen
  \bibfield  {author} {\bibinfo {author} {\bibfnamefont {T.}~\bibnamefont
  {Hikihara}}\ and\ \bibinfo {author} {\bibfnamefont {T.}~\bibnamefont
  {Suzuki}},\ }\bibfield  {title} {\bibinfo {title} {{Long-distance
  entanglement in one-dimensional quantum systems under sinusoidal
  deformation}},\ }\href {https://link.aps.org/doi/10.1103/PhysRevA.87.042337}
  {\bibfield  {journal} {\bibinfo  {journal} {Phys. Rev. A}\ }\textbf {\bibinfo
  {volume} {87}},\ \bibinfo {pages} {042337} (\bibinfo {year}
  {2013})}\BibitemShut {NoStop}%
\bibitem [{\citenamefont {Nishimoto}\ \emph {et~al.}(2013)\citenamefont
  {Nishimoto}, \citenamefont {Shibata},\ and\ \citenamefont
  {Hotta}}]{nishimoto2013ncom}%
  \BibitemOpen
  \bibfield  {author} {\bibinfo {author} {\bibfnamefont {S.}~\bibnamefont
  {Nishimoto}}, \bibinfo {author} {\bibfnamefont {N.}~\bibnamefont {Shibata}},\
  and\ \bibinfo {author} {\bibfnamefont {C.}~\bibnamefont {Hotta}},\ }\bibfield
   {title} {\bibinfo {title} {{Controlling frustrated liquids and solids with
  an applied field in a kagome Heisenberg antiferromagnet}},\ }\href
  {https://doi.org/10.1038/ncomms3287} {\bibfield  {journal} {\bibinfo
  {journal} {Nat. Commun.}\ }\textbf {\bibinfo {volume} {4}},\ \bibinfo {pages}
  {2287} (\bibinfo {year} {2013})}\BibitemShut {NoStop}%
\bibitem [{\citenamefont {Ichioka}\ \emph {et~al.}(2001)\citenamefont
  {Ichioka}, \citenamefont {Kaneshita},\ and\ \citenamefont
  {Machida}}]{ichioka2001jpsj}%
  \BibitemOpen
  \bibfield  {author} {\bibinfo {author} {\bibfnamefont {M.}~\bibnamefont
  {Ichioka}}, \bibinfo {author} {\bibfnamefont {E.}~\bibnamefont {Kaneshita}},\
  and\ \bibinfo {author} {\bibfnamefont {K.}~\bibnamefont {Machida}},\
  }\bibfield  {title} {\bibinfo {title} {{Collective Modes of Incommensurate
  Spin Density Wave in One-Dimensional Hubbard Model}},\ }\href
  {https://doi.org/10.1143/JPSJ.70.818} {\bibfield  {journal} {\bibinfo
  {journal} {J. Phys. Soc. Jpn.}\ }\textbf {\bibinfo {volume} {70}},\ \bibinfo
  {pages} {818} (\bibinfo {year} {2001})}\BibitemShut {NoStop}%
\bibitem [{\citenamefont {Ono}\ and\ \citenamefont
  {Hamano}(2000)}]{ono2000jpsj}%
  \BibitemOpen
  \bibfield  {author} {\bibinfo {author} {\bibfnamefont {Y.}~\bibnamefont
  {Ono}}\ and\ \bibinfo {author} {\bibfnamefont {T.}~\bibnamefont {Hamano}},\
  }\bibfield  {title} {\bibinfo {title} {Peierls distortion in two-dimensional
  tight-binding model},\ }\href@noop {} {\bibfield  {journal} {\bibinfo
  {journal} {J. Phys. Soc. Jpn.}\ }\textbf {\bibinfo {volume} {69}},\ \bibinfo
  {pages} {1769} (\bibinfo {year} {2000})}\BibitemShut {NoStop}%
\bibitem [{\citenamefont {Bethe}(1931)}]{bethe1931zp}%
  \BibitemOpen
  \bibfield  {author} {\bibinfo {author} {\bibfnamefont {H.~A.}\ \bibnamefont
  {Bethe}},\ }\href@noop {} {\bibfield  {journal} {\bibinfo  {journal} {Z.
  Phys.}\ }\textbf {\bibinfo {volume} {71}},\ \bibinfo {pages} {205} (\bibinfo
  {year} {1931})}\BibitemShut {NoStop}%
\bibitem [{\citenamefont {Lieb}\ and\ \citenamefont {Wu}(1968)}]{lieb1968prl}%
  \BibitemOpen
  \bibfield  {author} {\bibinfo {author} {\bibfnamefont {E.~H.}\ \bibnamefont
  {Lieb}}\ and\ \bibinfo {author} {\bibfnamefont {F.~Y.}\ \bibnamefont {Wu}},\
  }\bibfield  {title} {\bibinfo {title} {{Absence of Mott Transition in an
  Exact Solution of the Short-Range, One-Band Model in One Dimension}},\ }\href
  {https://link.aps.org/doi/10.1103/PhysRevLett.20.1445} {\bibfield  {journal}
  {\bibinfo  {journal} {Phys. Rev. Lett.}\ }\textbf {\bibinfo {volume} {20}},\
  \bibinfo {pages} {1445} (\bibinfo {year} {1968})}\BibitemShut {NoStop}%
\bibitem [{\citenamefont {Bernevig}\ \emph {et~al.}(2006)\citenamefont
  {Bernevig}, \citenamefont {Orenstein},\ and\ \citenamefont
  {Zhang}}]{bernevig2006prl}%
  \BibitemOpen
  \bibfield  {author} {\bibinfo {author} {\bibfnamefont {B.~A.}\ \bibnamefont
  {Bernevig}}, \bibinfo {author} {\bibfnamefont {J.}~\bibnamefont
  {Orenstein}},\ and\ \bibinfo {author} {\bibfnamefont {S.-C.}\ \bibnamefont
  {Zhang}},\ }\bibfield  {title} {\bibinfo {title} {{Exact SU(2) Symmetry and
  Persistent Spin Helix in a Spin-Orbit Coupled System}},\ }\href
  {https://link.aps.org/doi/10.1103/PhysRevLett.97.236601} {\bibfield
  {journal} {\bibinfo  {journal} {Phys. Rev. Lett.}\ }\textbf {\bibinfo
  {volume} {97}},\ \bibinfo {pages} {236601} (\bibinfo {year}
  {2006})}\BibitemShut {NoStop}%
\bibitem [{\citenamefont {Schliemann}(2017)}]{john2017rmp}%
  \BibitemOpen
  \bibfield  {author} {\bibinfo {author} {\bibfnamefont {J.}~\bibnamefont
  {Schliemann}},\ }\bibfield  {title} {\bibinfo {title} {{Colloquium:
  Persistent spin textures in semiconductor nanostructures}},\ }\href
  {https://link.aps.org/doi/10.1103/RevModPhys.89.011001} {\bibfield  {journal}
  {\bibinfo  {journal} {Rev. Mod. Phys.}\ }\textbf {\bibinfo {volume} {89}},\
  \bibinfo {pages} {011001} (\bibinfo {year} {2017})}\BibitemShut {NoStop}%
\bibitem [{\citenamefont {Park}\ \emph {et~al.}(2020)\citenamefont {Park},
  \citenamefont {Yang},\ and\ \citenamefont {Lee}}]{park2020prr}%
  \BibitemOpen
  \bibfield  {author} {\bibinfo {author} {\bibfnamefont {H.~K.}\ \bibnamefont
  {Park}}, \bibinfo {author} {\bibfnamefont {H.-J.}\ \bibnamefont {Yang}},\
  and\ \bibinfo {author} {\bibfnamefont {S.}~\bibnamefont {Lee}},\ }\bibfield
  {title} {\bibinfo {title} {{Spin-helix-driven insulating phase in
  two-dimensional lattice}},\ }\href
  {https://link.aps.org/doi/10.1103/PhysRevResearch.2.033487} {\bibfield
  {journal} {\bibinfo  {journal} {Phys. Rev. Research}\ }\textbf {\bibinfo
  {volume} {2}},\ \bibinfo {pages} {033487} (\bibinfo {year}
  {2020})}\BibitemShut {NoStop}%
\bibitem [{\citenamefont {Kubo}\ and\ \citenamefont
  {Kishi}(1990)}]{kubo1990prb}%
  \BibitemOpen
  \bibfield  {author} {\bibinfo {author} {\bibfnamefont {K.}~\bibnamefont
  {Kubo}}\ and\ \bibinfo {author} {\bibfnamefont {T.}~\bibnamefont {Kishi}},\
  }\bibfield  {title} {\bibinfo {title} {{Rigorous bounds on the
  susceptibilities of the Hubbard model}},\ }\href
  {https://link.aps.org/doi/10.1103/PhysRevB.41.4866} {\bibfield  {journal}
  {\bibinfo  {journal} {Phys. Rev. B}\ }\textbf {\bibinfo {volume} {41}},\
  \bibinfo {pages} {4866} (\bibinfo {year} {1990})}\BibitemShut {NoStop}%
\bibitem [{\citenamefont {Furukawa}\ and\ \citenamefont
  {Imada}(1991)}]{furukawa1991jpsj}%
  \BibitemOpen
  \bibfield  {author} {\bibinfo {author} {\bibfnamefont {N.}~\bibnamefont
  {Furukawa}}\ and\ \bibinfo {author} {\bibfnamefont {M.}~\bibnamefont
  {Imada}},\ }\bibfield  {title} {\bibinfo {title} {{Charge Gap, Charge
  Susceptibility and Spin Correlationin the Hubbard Model on a Square
  Lattice}},\ }\href {https://doi.org/10.1143/JPSJ.60.3604} {\bibfield
  {journal} {\bibinfo  {journal} {J. Phys. Soc. Jpn.}\ }\textbf {\bibinfo
  {volume} {60}},\ \bibinfo {pages} {3604} (\bibinfo {year}
  {1991})}\BibitemShut {NoStop}%
\bibitem [{\citenamefont {Assaad}\ and\ \citenamefont
  {Imada}(1996)}]{assaad1996prl}%
  \BibitemOpen
  \bibfield  {author} {\bibinfo {author} {\bibfnamefont {F.~F.}\ \bibnamefont
  {Assaad}}\ and\ \bibinfo {author} {\bibfnamefont {M.}~\bibnamefont {Imada}},\
  }\bibfield  {title} {\bibinfo {title} {{Insulator-Metal Transition in the
  One- and Two-Dimensional Hubbard Models}},\ }\href
  {https://link.aps.org/doi/10.1103/PhysRevLett.76.3176} {\bibfield  {journal}
  {\bibinfo  {journal} {Phys. Rev. Lett.}\ }\textbf {\bibinfo {volume} {76}},\
  \bibinfo {pages} {3176} (\bibinfo {year} {1996})}\BibitemShut {NoStop}%
\end{thebibliography}%
\end{document}